\documentclass[a4paper,fleqn,usenatbib]{mnras}

\usepackage[T1]{fontenc}
\usepackage{ae,aecompl}

\usepackage{graphicx}	
\usepackage{pdflscape}
\usepackage{epstopdf}
\usepackage{amsmath}	
\usepackage{amssymb}	
\usepackage{booktabs}

\usepackage[dvipsnames]{xcolor} 
\usepackage{color,soul} 
\usepackage[normalem]{ulem}
\usepackage{subfig,graphicx}

\newcommand\blfootnote[1]{%
  \begingroup
  \renewcommand\thefootnote{}\footnote{#1}%
  \addtocounter{footnote}{-1}%
  \endgroup
}

\title[]{A Kinematical Study of the Dwarf Irregular Galaxy NGC\,1569 and its Supernova Remnants}

\author[M. S\'anchez-Cruces]{
M. S\'anchez-Cruces$^{1}$,\thanks{msanchez@astro.unam.mx: }
M.M. Sardaneta$^{2}$,
I. Fuentes-Carrera$^{3}$,
M. Rosado$^{1}$,
\newauthor{N. C\'ardenas-Mart\'inez $^{3}$},
and M. A. Lara-López$^{4,5,6}$ \\
$^{1}$ Instituto de Astronom\'ia, Universidad Nacional Aut\'onoma de M\'exico, Circuito Exterior, C.U., Apartado Postal 70-264, 04510 \\
Ciudad de M\'exico, M\'exico\\
$^{2}$ Aix Marseille Univ, CNRS, CNES, LAM, Laboratoire d’Astrophysique de Marseille, Marseille, France\\
$^{3}$ Escuela Superior de F\'isica y Matem\'aticas, Instituto Polit\'ecnico Nacional, U.P. Adolfo L\'opez Mateos, C.P. 07738,\\
Ciudad de M\'exico, M\'exico\\
$^{4}$ Armagh Observatory and Planetarium, College Hill, Armagh BT61 9DG, Northern Ireland, UK \\
$^{5}$ Departamento de Física de la Tierra y Astrofísica, Universidad Complutense de Madrid, E-28040 Madrid, Spain\\
$^{6}$ Instituto de Física de Partículas y del Cosmos IPARCOS, Fac. de Ciencias Físicas, Universidad Complutense de Madrid, \\
E-28040, Madrid, Spain}
\date{Accepted XXX. Received YYY; in original form ZZZ}

\pubyear{2022}

\begin{document}
\label{firstpage}
\maketitle

\begin{abstract}

We present Fabry-Pérot observations in the  H$\alpha$ and [{S\,{\sc ii}}] lines to study the kinematics of the Magellanic-type dwarf irregular galaxy NGC\,1569, these observations allowed us to computed the H$\alpha$ velocity field of this galaxy. 
Doing a detailed analysis of the velocity along the line-of-sight and H$\alpha$ velocity profiles, we identified the origin of most of the motions in the innermost parts of the galaxy and discarded the possibility of deriving a rotation curve that traces the gravitational well of the galaxy.
We analysed the kinematics of the ionised gas around 31 supernova remnants previously detected in NGC\,1569 by other authors, in optical and radio emission. We found that the H$\alpha$ velocity profiles of the supernova remnants are complex indicating the presence of shocks. Fitting these profiles with several Gaussian functions, we computed their expansion velocities which rank from 87 to 188 km s$^{-1}$ confirming they are supernova remnants. 
Also, we determined the physical properties such as  electron density, mechanical energy, and kinematic age for 30 of the 31 supernova remnants and found they are in the radiative phase with an energy range from 1 to 39 $\times$10$^{50}$ erg s$^{-1}$ and an age from  2.3 to 8.9$\times$10$^{4}$ yr.  Finally, we estimated the Surface Brightness - Diameter ($\Sigma$-D) Relation for NGC\,1569 and obtained a slope $\beta$ = 1.26$\pm$0.2, comparable with the $\beta$ value obtained for supernova remnants in galaxies M31 and M33.
 
\end{abstract}



\section{Introduction} \label{sec:intro}
Detection and identification of supernova remnants (SNRs) in nearby galaxies depend on multiple physical properties involving the interstellar medium (ISM) of the host galaxy, as well as the morphology and evolutionary stage determined by the supernova (SN) progenitor (Type Ia or core-collapse SN). By studying extragalactic SNRs we can determine some physical aspects that took place during their evolution and consequently, several key factors in understanding star formation activity, as well as, the formation and evolution of the host galaxy.

The first extragalactic SNRs  identifications were made in the Large Magellanic Cloud (LMC) by  \citet{Mathewson1963}, who  found three non-thermal radio sources. Additionally, \citet{Mathewson1972} \citet{Mathewson1973a}, \citet{Mathewson1973b} found 12 SNRs in the LMC and 2 in the Small Magellanic Cloud (SMC) using optical ([{S\,{\sc ii}}]/H$\alpha$ ratios>0.4) and radio emission imaging observations (using radio spectral indices $\alpha\leq$-0.2). Since then, research to identify extragalactic SNRs has been mainly done using radio, optical and X-ray emission  \citep[see for example,][]{Long1981,Seward-Mitchell1981,Georgelin1983, Kaastra-Mewe1995, AmbrocioCruz2006,AmbrocioCruz2017}.   

SNR radio emission is generally dominated by a non-thermal continuum. Since these objects are synchrotron sources their identification is usually achieved using their morphology and non-thermal radio spectral indices with values ranging from -0.2 to -0.7. X-ray emission of SNRs is dominated by thermal plasma heated by the SN explosion at temperatures of $\sim$10$^6$ K and  mainly emitting soft thermal X-rays (0.5 - 2.0 keV). Therefore, the identification in this waveband has been done first considering counterparts in other wavelengths along with their soft X-ray spectra \citep[see,][]{Pannuti2000, Nishiuchi2001, Long2010, Leonidaki2010, Wang1992, Wang1998}.  

The SNR optical emission is dominated by H$\alpha$ and forbidden lines, for example [{O\,{\sc iii}}]$\lambda\lambda$4959,5007~\AA, [{N\,{\sc ii}}]$\lambda\lambda$6549,6583~\AA, and [{S\,{\sc ii}}]$\lambda\lambda$6717,6731 \AA\ \citep[][and references therein]{Fesen1985}. 
The optical identification of SNRs is typically done using the emission lines ratio [{S\,{\sc ii}}]/H$\alpha$. In particular, this line ratio has been used to distinguish between SNRs with [{S\,{\sc ii}}]/H$\alpha$ ratios>0.4 and HII regions with [{S\,{\sc ii}}]/H$\alpha$ ratios<0.2 \citep[e.g.,][]{Mathewson1972, Mathewson1973b, Raymond1979, Dodorico1980, Long1990, Levenson1995, Blair1997, Matonick1997}.  It is important to take into account that the  [{S\,{\sc ii}}]/H$\alpha$>0.4 value is valid for galaxies with low metallicities such as the LMC with Z$_{LMC}\sim$0.3 -0.5 Z$\odot$ \citep{Russell-Dopita992} and the SMC with Z$_{SMC}\sim$0.2 Z$\odot$ \citep{Russell-Dopita992}; while for galaxies with similar metallicities to the Milky Way, this value is about a unit, due to the [{S\,{\sc ii}}] line being comparable with the H$\alpha$ line \citep[see also][]{Dodorico1976}.

Being this the most common method of detection, recent observations and surveys have identified a large sample of SNRs in nearby galaxies using the classical definition for optical detection of extragalactic SNRs ([{S\,{\sc ii}}]/H$\alpha$ ratios$>$0.4)  \citep[e.g.,][]{Vucetic2015, Winkler2017,Long2018,Long2019,Moumen2019}. 
However, in some cases the SNR's physical properties such as electronic density, kinetic energy, kinematic age and possible evolutionary stage, remain unknown due to the lack of spatial and spectral resolution. 

Another method to identify and study the physical properties of SNRs is the kinematical method proposed and used to detect SNRs in the Magellanic Clouds by \citet{Rosado1993,Rosado1993a}. This method consists on the detection of the suspected nebula in several velocity channels of a scanning high resolution Fabry-Pérot (FP) interferometer data cube based on the fact that the SNRs radial velocity profiles are complex, i.e., presenting velocity splitting, high velocity wings or humps \citep{Rosado1993a}. These features can indicate the presence of more than one velocity component due to the velocity of the SNR expanding shell. The characteristic widths of about 200~km~s$^{-1}$ are also due to the presence of shocks. The characteristic expansion velocity of these complex velocity profiles is given as follows: the difference between the largest negative value of the velocity in the profile and the largest positive velocity value.

In general, for HII regions the velocity profiles are fitted with a simple component, while the SNRs complex velocity profiles can be fitted with multiple velocity components \citep[see][]{LeCoarer1993}.
Thus, the kinematical method is useful in the identification of SNRs inside bright HII regions and superbubbles\footnote{Superbubbles are objects blown by fast stellar winds and supernova explosions from groups of massive stars \citep{Tomisaka1981, MacLow1988, Chu2008} with a ring shaped nebulae formed inside the disc of the galaxy with sizes of the order of 10 - 100 pc.} formed by previous wind driven nebula.

One way to see intrinsic broadening of the emission lines and distinguish between shock-ionised (i.e., SNRs) or photo-ionised (i.e., HII regions) regions is by using high spectroscopic resolution observations \citep[see,][]{Points2019}; another way is by using data from an integral field spectrograph as the FP interferometer, from which we can also get kinematic information of each pixel over all the field of view (FoV).

With high resolution FP interferometry data it is possible to study the kinematics of Galactic and extragalactic SNRs obtaining the SNR expansion velocity, the energy deposited in the ISM by the SN explosion and the age of the SNR considering its evolutionary stage  \citep[see,][]{Rosado1981, Rosado1982, Rosado1996, AmbrocioCruz2014, Sanchez-Cruces2018}. In addition, FP data allows us to study the global kinematics of galaxies \citep{FuentesCarrera2019, GomezLopez2019, Sardaneta2020} and their regions with particular physical features such as perturbations induced by interactions \citep{FuentesCarrera2004, Repetto2010} or regions with a high star formation rate  \citep[SFR, see][]{CardenasMartinez2018}. Thus, the kinematical method complements the classic optical method of SNR detection, as well as radio and X-ray observations by helping identify SNRs inside bright HII regions.

The determination of the kinematics of dwarf irregular galaxies is a powerful tool to estimate the total matter distribution at small scales.
In particular, HI observations have allowed the determination of rotation curves (RCs) for this type of galaxies, since neutral hydrogen is typically detected well beyond the optical disc and is not affected by dust extinction \citep{Reakes1980, Stil2002, Johnson2012, Lelli2014, Iorio2017}.
However, the angular resolution of HI observations is not very large.
Combining the RCs obtained for central regions of galaxies from ionised gas data with the results from neutral hydrogen data for the more external regions is becoming more frequent \citep{Moiseev2014, Egorov2021}.
On the other hand, it is rather difficult to measure the RC of gas in dwarf galaxies. 
However, due to the shallow potential well of dwarf irregular galaxies, the interaction of young stellar groups with the ISM in this type of galaxies can be particularly important. 
The ionising radiation of OB stars, as well as the kinetic energy of stellar winds and supernova explosions can heat the gas, forming bubbles, shells, outflows and chaotic turbulent motions in the gaseous disc. 
In some cases, these motions dominate the kinematics and dynamics in the inner parts of dwarf galaxies, so a reliable RC that properly describes or traces the potential well of the galaxy cannot be derived from the ionised gas as suggested by \citet{Moiseev2012, Moiseev2015, Egorov2021}.

\subsection{SNRs in NGC 1569}
\label{SNR} 
The Magellanic-type dwarf irregular galaxy (dIm) NGC\,1569, a well known starburst
galaxy, lies at a distance of 3.36 Mpc \citep[][ measured using the tip of the red giant branch (TRGB) with Hubble Space Telescope (HST) observations]{Grocholski2008} and presents a low  metallicity \citep[Z = 0.25 Z$\odot$,][]{Devost1997, Kobulnicky-Skillman1997} which lies between the metallicities of the SMC and LMC. Large-scale shells and superbubbles are very well defined in H$\alpha$ emission \citep{Hunter1993, Westmoquette2008} and coincide with extended X-ray emission \citep{Martin2002, Sanchez-Cruces2015}. Using optical observations in H$\alpha$ and [SII] carried out with the UNAM Scanning Fabry-Pérot Interferometer PUMA, \citet{Sanchez-Cruces2015} detected several superbubbles as well as filamentary structure and supershells in this galaxy, some of them found previously by \citet{Hunter1993} and \citet{Westmoquette2008}.

\label{F1}
\begin{figure*}\centering
\includegraphics[width=2\columnwidth]{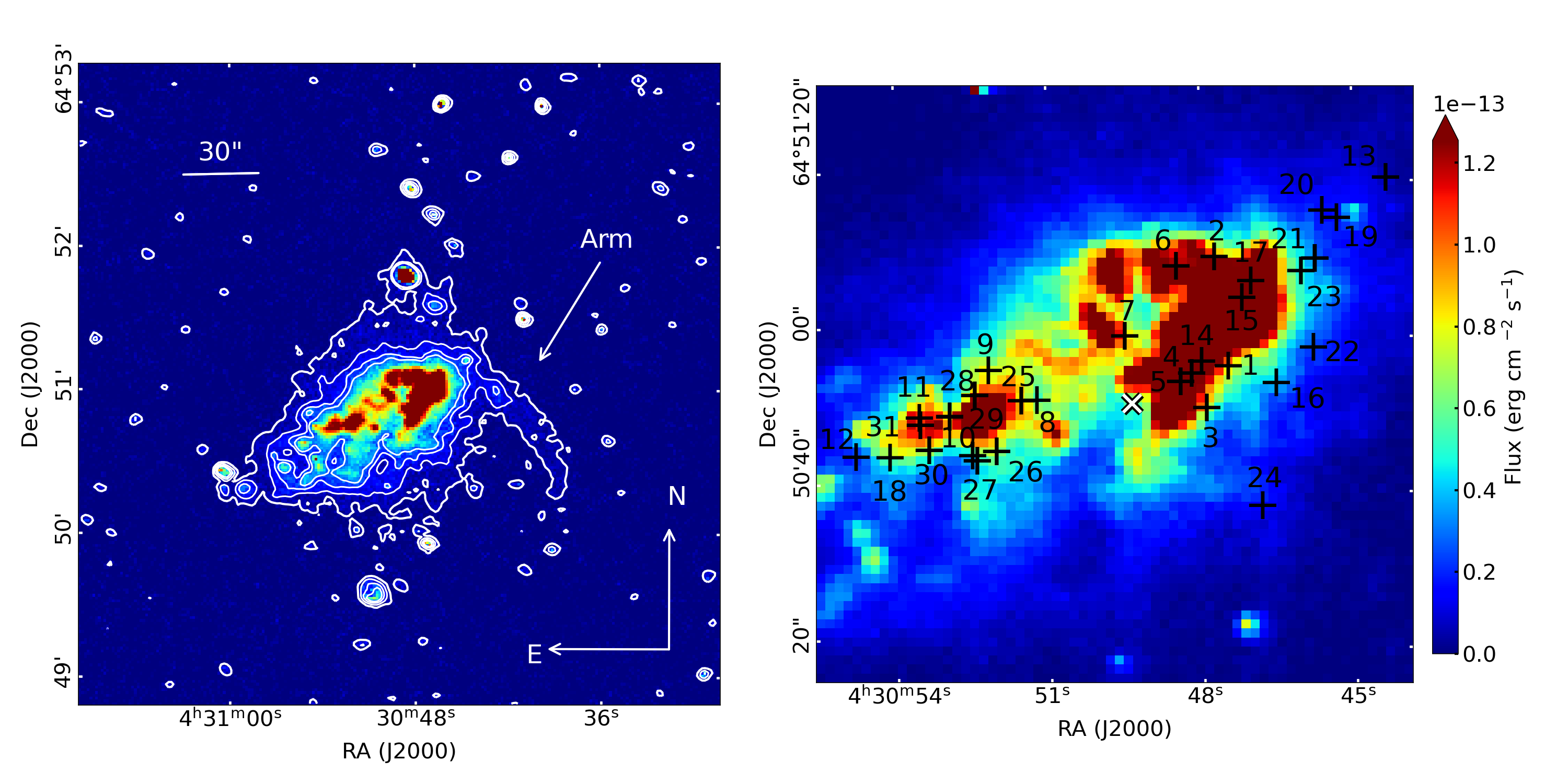}
  \caption{\textit{Left}: Direct image of NGC\,1569 in H$\alpha$ obtained with PUMA \citep[from][]{Sanchez-Cruces2015}. The most external contour is at 1.03 $\times$10$^{-14}$ erg s$^{-1}$. Next three contours at 3.69, 7.58 and 11.04 $\times$10$^{-14}$ erg s$^{-1}$ are from  \citet{Sanchez-Cruces2015}. \textit{Right}: NGC\,1569 disc close-up  in H$\alpha$ emission. Crosses indicate the SNRs positions classified by \citet{Greve2002}, \citet{Chomiuk2009} and \citet{Ercan2018}. The white `x' indicates the position of the kinematical centre obtained in this work (see Section \ref{CR}).}
  \label{Ha_snr}
\end{figure*}

For this galaxy, \citet{Greve2002} detected five possible radio SNRs as non-thermal radio sources located in the superstar clusters region using the Multi Element Radio Linked Interferometry Network (MERLIN) observations at 1.4 and 5 GHz. 

\citet{Chomiuk2009} detected 23 SNRs candidates in NGC\,1569 using the Very Large Array (VLA) radio continuum data at 20, 6, and 3.6 cm, as part of a SNRs survey in four nearby irregular galaxies. The radio continuum sample obtained by those authors is conformed by SNRs, HII regions and background radio galaxies; therefore, to differentiate SNRs from HII regions they used the radio spectral indices considering $\alpha\leq$-0.2 for SNRs and $\alpha>$-0.2 for HII regions. Then, to differentiate detected SNRs from background galaxies, they considered that background galaxies did not have H$\alpha$ emission. Therefore, the SNRs sample obtained by  \citet{Chomiuk2009} was done considering radio continuum and optical emission.

Recently, \citet{Ercan2018} showed evidence of  13 possible optical SNRs candidates in this galaxy with CCD imaging using the [{S\,{\sc ii}}]/H$\alpha\geq$ 0.4 standard criteria. They identified the new SNRs candidates and computed the electron density (ne), of about 121$\pm$17 cm$^{-3}$ for only one SNR candidate using the [{S\,{\sc ii}}]$\lambda$6716/[{S\,{\sc ii}}]$\lambda$6731 average flux ratio. Also, with long-slit spectroscopic observations, they found that the shock wave velocity (Vs) of one of the 13 SNRs ranges from 100 to 150 km s$^{-1}$ using the [OIII]$\lambda$5007/H$\beta$ ratio. Finally, those authors used Chandra data to extract the X-ray spectra of the 13 SNRs candidates; their results indicate that only 10 of the 13 SNRs candidates had a spectrum in the 0.5-2.0 keV band with good statistics, confirming the existence of 10 SNRs candidates in that wavelength.

Taking into account the three SNRs samples in radio and optical emission reported previously, this galaxy harbours a total of 31 SNRs listed in Table \ref{List_SNR} and pointed out in a H$\alpha$ image in the right panel of Figure \ref{Ha_snr}. 
 
In \citet{Sanchez-Cruces2015}, hereafter Paper~I, we studied the kinematics of superbubbles and supershells in NGC\,1569 and focused on the nature and formation of those objects.
In this work, we studied the global kinematics of this irregular galaxy and used the kinematic method to verify the optical SNRs hosted in NGC\,1569 found by other authors. We used a distance to NGC\,1569 of 3.36 Mpc as indicated above. In Section 2 we present the observations and data reduction. In Section 3 the global kinematics of the galaxy is presented. In Section 4 we show the kinematical analysis of the SNRs. Finally, the conclusions are presented in Section 5.

\begin{table*}\centering
  \setlength{\tabcolsep}{\tabcolsep}
  \caption{ Supernova remnants of NGC 1569.}
  \label{List_SNR}
  \label{Table1}
  \scalebox{1}{
 \begin{tabular}{lccccccccccccccc}
    \hline
ID & 
\multicolumn{1}{c}{RA}&
\multicolumn{1}{c}{Dec.}&
\multicolumn{1}{c}{Maj/Min Axis	}&	
\multicolumn{1}{c}{Maj/Min Axis$^*$}&
\multicolumn{1}{c}{Previous}\\
&
\multicolumn{1}{c}{($^{h}$:$^{m}$:$^{s}$)}&
\multicolumn{1}{c}{($\degr$:$\arcmin$:$\arcsec$)}&
\multicolumn{1}{c}{(arcsec)}&
\multicolumn{1}{c}{(pc)}&
\multicolumn{1}{c}{identification}\\
\hline
SNR1	&	04:30:47.48	&	64:50:55.70	&	2.1/1.7$^{a}$	&	34.2/27.7	&	SNR1$^{c}$,	N1569-16$^{d}$	\\
SNR2	&	04:30:47.73	&	64:51:09.70	&	2.3/1.8$^{a}$	&	37.5/29.3	&	SNR2$^{c}$,	N1569-17$^{d}$	\\
SNR3	&	04:30:47.91	&	64:50:50.30	&	2.9/1.5$^{a}$	&	47.2/24.4	&	SNR3$^{c}$,	N1569-18$^{d}$	\\
SNR4	&	04:30:48.20	&	64:50:54.70	&	1.5/1.3$^{a}$	&	24.4/21.2	&	SNR4$^{c}$,	N1569-20$^{d}$	\\
SNR5	&	04:30:48.42	&	64:50:53.60	&	2.7/2.3$^{a}$	&	44.0/37.5	&	SNR5$^{c}$,	N1569-21$^{d}$	\\
SNR6	&	04:30:48.48	&	64:51:08.50	&	1.8/1.6$^{a}$	&	29.3/26.1	&	SNR6$^{c}$,	N1569-23$^{d}$	\\
SNR7	&	04:30:49.50	&	64:50:59.40	&	3.2/2.8$^{a}$	&	52.1/45.6	&	SNR7$^{c}$,	N1569-27$^{d}$	\\
SNR8	&	04:30:51.24	&	64:50:51.05	&	2.1$^{b}$	&	34.2	&	SNR8$^{c}$	\\	
SNR9	&	04:30:52.19	&	64:50:54.80	&	3.4/2.2$^{a}$	&	55.4/35.8	&	SNR9$^{c}$,	N1569-32$^{d}$	\\
SNR10	&	04:30:52.51	&	64:50:43.86	&	2.1$^{b}$	&	34.2	&	SNR10$^{c}$	\\	
SNR11	&	04:30:53.55	&	64:50:48.68	&	2.1$^{b}$	&	34.2	&	SNR11$^{c}$	\\	
SNR12	&	04:30:54.80	&	64:50:43.50	&	2.1/1.9$^{a}$	&	34.2/31.0	&	SNR12$^{c}$,	N1569-38$^{d}$	\\
SNR13	&	04:30:44.35	&	64:51:20.01	&	1.7/1.5$^{a}$	&	27.7/24.4	&	SNR13$^{c}$,	N1569-4$^{d}$	\\
SNR14	&	04:30:48.00	&	64:50:56.30	&	2.1$^{b}$	&	34.2	&	VLA-8$^{f}$	\\	
SNR15	&	04:30:47.20	&	64:51:04.50	&	2.1$^{b}$	&	34.2	&	VLA-6$^{f}$	\\	
SNR16	&	04:30:46.51	&	64:50:53.40	&	1.6/1.4$^{a}$	&	26.1/22.8	&	M2$^{f}$,	N1569-11$^{d}$	\\
SNR17	&	04:30:47.02	&	64:51:06.70	&	1.5/1.4$^{a}$	&	24.4/22.8	&	M3$^{f}$,	N1569-14$^{d}$	\\
SNR18	&	04:30:54.13	&	64:50:43.50	&	2.1/1.9$^{a}$	&	34.2/31.0	&	M6/VLA-1$^{f}$,	N1569-38$^{d}$	\\
SNR19	&	04:30:45.32	&	64:51:14.90	&	4.9/2.0$^{a}$	&	79.8/32.6	&	N1569-05$^{d}$	\\	
SNR20	&	04:30:45.60	&	64:51:15.80	&	1.5/1.4$^{a}$	&	24.4/22.8	&	N1569-06$^{d}$	\\	
SNR21	&	04:30:45.75	&	64:51:09.60	&	1.6/1.4$^{a}$	&	26.1/22.8	&	N1569-07$^{d}$	\\	
SNR22	&	04:30:45.79	&	64:50:58.20	&	1.7/1.6$^{a}$	&	27.7/26.1	&	N1569-08$^{d}$	\\	
SNR23	&	04:30:46.03	&	64:51:08.00	&	1.6/1.5$^{a}$	&	26.1/24.4	&	N1569-09$^{d}$	\\	
SNR24	&	04:30:46.83	&	64:50:37.80	&	2.3/2.0$^{a}$	&	37.5/32.6	&	N1569-12$^{d}$	\\	
SNR25	&	04:30:51.55	&	64:50:51.00	&	1.7/1.4$^{a}$	&	27.7/22.8	&	N1569-28$^{d}$	\\	
SNR26	&	04:30:52.04	&	64:50:44.40	&	1.8/1.4$^{a}$	&	29.3/22.8	&	N1569-30$^{d}$	\\	
SNR27	&	04:30:52.42	&	64:50:43.20	&	1.7/1.5$^{a}$	&	27.7/24.4	&	N1569-33$^{d}$	\\	
SNR28	&	04:30:52.46	&	64:50:51.60	&	1.9/1.4$^{a}$	&	31.0/22.8	&	N1569-34$^{d}$	\\	
SNR29	&	04:30:52.96	&	64:50:48.80	&	2.1/1.5$^{a}$	&	34.2/24.4	&	N1569-35$^{d}$	\\	
SNR30	&	04:30:53.36	&	64:50:44.50	&	1.7/1.4$^{a}$	&	27.7/22.8	&	N1569-36$^{d}$	\\	
SNR31	&	04:30:53.53	&	64:50:47.70	&	3.2/1.6$^{a}$	&	52.1/26.1	&	N1569-37$^{d}$	\\	
\hline
\end{tabular}}\\
\begin{flushleft}
The columns are as follows:\\
Column 1 : SNR identification\\
Column 2-3 : SNR right ascension and declination (epoch 2000).\\
Column 4 : Major and minor axis size in arcsec.$^a$ Taken from \citet{Chomiuk2009} in radio emission. $^b$ Average radius of SNRs.\\
Column 5 : Major and minor axis size in pc. $^*$ Calculated in this work using a distance to the galaxy of D=3.36 Mpc.\\
Column 6 : Previous identifications. $^c$\citet{Ercan2018}. $^d$\citet{Chomiuk2009}. $^f$\citet{Greve2002}.\\
\end{flushleft}
\end{table*}

\section{Observations and Data Reduction} 
\subsection{Observations}
\label{observations}
The observational data used in this work are the same as in Paper~I, and were obtained in November 1997 using the 2.1 m telescope of the Observatorio Astron\'omico Nacional of the Universidad Nacional Aut\'onoma de M\'exico (OAN-UNAM), at San Pedro M\'artir, B. C., M\'exico with the UNAM scanning Fabry-Pérot interferometer PUMA \citep{Rosado1995}. No additional FP observations of NGC\,1569 have been obtained recently for this work. 

We used a 1024 $\times$ 1024 Tektronix CCD detector binned by a factor of 2, resulting in an image size of 512 $\times$ 512 px and a plate scale of 1.18 arcsec/pixel (44 pc/pixel). The FoV of PUMA is 10' and the spectral resolution is 0.41 \AA\ (equivalent to a velocity resolution of 19.0 km s$^{-1}$) at H$\alpha$. More details about the instrument's main characteristics are given in Paper I.

We  obtained a set of direct images in H$\alpha$ and [{S\,{\sc ii}}]$\lambda\lambda$6717,6731 \AA\ lines using PUMA in its direct-imaging mode (i.e., images taken with the FP etalon out of the telescope's line-of-sight), using the two-binning format and the interference filters centred at 6570 \AA\ for H$\alpha$ and 6720 \AA\ for [{S\,{\sc ii}}], both with a bandwidth of 20 \AA\  \citep[see Paper I][]{Sanchez-Cruces2015}.  Then, with each filter, we scanned the  Fabry-Pérot interferometer PUMA through 48 channels getting two  object data cubes of dimensions  512 $\times$ 512 $\times$ 48.  For both H$\alpha$ and [{S\,{\sc ii}}] data cubes, the integration time was 60 s per channel. To calibrate both data cubes, we obtained a calibration data cube using a Hydrogen lamp (6562.78 \AA\ wavelength calibration) for the  H$\alpha$ cube and a Neon lamp (6598.95 \AA\ wavelength calibration) for the [{S\,{\sc ii}}] cube. These calibration data cubes are the same dimensions as the object data cubes.

\subsection{Data Reduction} 
The reduction and analysis of Fabry-Pérot data were performed using the ADHOCw\footnote{\url{http://cesam.lam.fr/fabryperot/index/softwares} developed by J. Boulesteix.} software. Both softwares, ADHOCw and CIGALE (used in Paper I), allowed us first make standard corrections (removal of cosmic rays and bias subtraction), second, to compute a wavelength-calibrated data cube, and lastly, to obtain monochromatic and continuum images, as well as, velocity and velocity dispersion maps.

To compute the wavelength-calibrated data cube, ADHOCw uses the calibration data cube to obtain the phase map. This is a 2D map that provides the reference wavelength for the line profile observed for each pixel.  When we apply the phase map to the observed data cube, the wavelength-calibrated data cube is created, which is also known as the velocity or wavelength data cube. Therefore, we got a wavelength data cube for both [{S\,{\sc ii}}]$\lambda\lambda$6717,6731 and H$\alpha$ emission lines by using their respective calibration cubes.

In order to improve the  signal-to-noise ratio in the outer parts of the galaxy, we applied two types of Gaussian smoothing on the wavelength data cube using ADHOCw software: a spectral smoothing with FWHM of 2 channels ($\sigma$=38~km~s$^{-1}$) and a spatial smoothing (x,y) with FWHM of (2,2) pixels. 

Spectral smoothing is done over the channels multiplying by a factor that depends on the amplitude of the Gaussian smoothing selected. Once this is done, the final spectral resolution is $\sim$0.60 \AA\ ($\sim$28.0 km s$^{-1}$). Although spatial smoothing increases the signal-to-noise ratio of the image, spatial resolution is lost, going from 1.18$\arcsec$ per pixel to $\sim$1.8$\arcsec$ per pixel.

The monochromatic image was obtained by measuring pixel by pixel the intensity inside the line above the local continuum (over 0.7 of the intensity peak fraction). The line-of-sight velocity\footnote{In this work we shall use the expression ``line-of-sight velocity'' or ``velocity along the line-of-sight'', $V_{LOS}$, instead of the expression ``radial velocity'' commonly used in the literature, in order to avoid confusion with the radial component of the velocity in the frame of reference of the disc of the galaxy.} field was obtained by computing the barycenter of the velocity profile for each pixel. 
From the velocity profiles of each pixel, ADHOCw computes its FWHM creating the observed velocity dispersion map ($\sigma_{obs}$). After correcting for instrumental ($\sigma_{inst}^{2}$) and thermal ($\sigma_{th}^{2}$) broadening, we obtained the velocity dispersion ($\sigma_{corr}$) according to: $\sigma_{corr}$ = $\sqrt{\sigma_{obs}^{2} - \sigma_{inst}^{2} - \sigma_{th}^{2}}$. The instrumental broadening $\sigma_{inst}$  was computed from the deconvolution of the Airy function and in this case corresponds to 38~km~s$^{-1}$. We computed the thermal broadening according to $\sigma_{th}$ = $\sqrt{82.5(\rm T_4/A)}$ where T$_4$ = T/10$^{4}$~K  and A is the atomic weight of the atom, so that $\sigma_{th}$ = 9.1 km s$^{-1}$ for hydrogen gas at T$_e$=10$^{4}$~K \citep{Spitzer1978}.

In order to quantify the accuracy of the velocity values determined through the barycenter method, we computed the barycenter position of the calibration data cube emission line for each pixel. We thus created the velocity map of the calibration, providing proof that the velocity accuracy for our data is about $\pm$2~km~s$^{-1}$.

\label{F2}
\begin{figure*}\centering
\includegraphics[width=2\columnwidth]{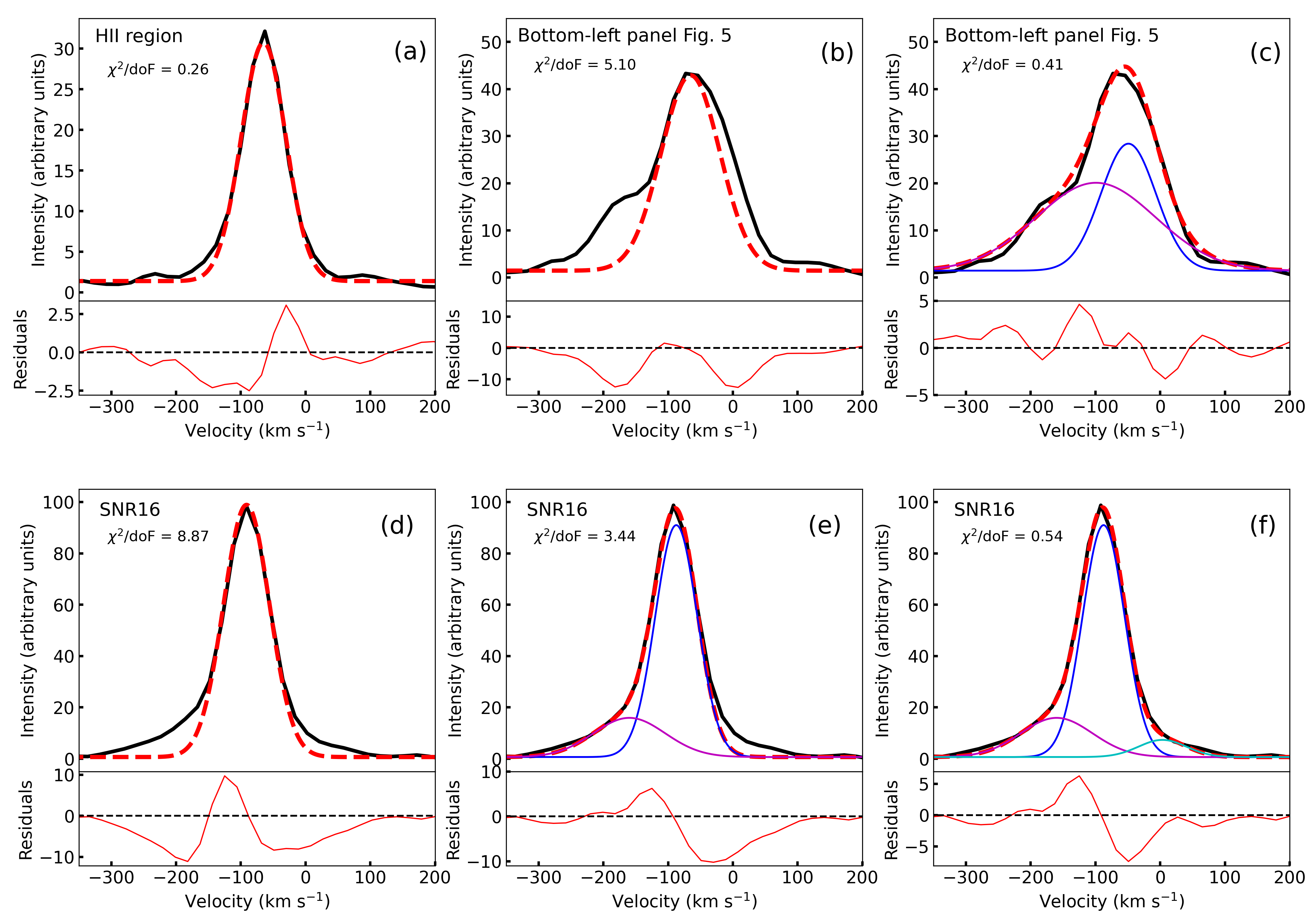}
  \caption{Examples of representative observed H$\alpha$ velocity profiles in NGC\,1569 (solid black line) fitted from different object or regions within NGC\,1569.  {\it Panel (a)}: Velocity profile of HII region no.12 classified by \citet {Waller1991}. The red dashed line is the velocity profile fit.   {\it  Panel (b)}: Velocity profile of a pixel located in central part of the galaxy (see Figure \ref{Central} in Section \ref{VF}).  Velocity profile fitted with a single Gaussian component ($\chi^2$/doF=5.10), which is not in agreement with the observed velocity profile. A better fit is shown in {\it Panel (c)} with a $\chi^2$/doF=0.41, which was performed with two velocity components highlighted in blue and magenta. Bottom panels show the fit process of the SNR16 velocity profile (see Section \ref{SNR_param}). {\it Panel (d)}: Velocity profile fitted with a single Gaussian fit  ($\chi^2$/doF=8.87), which is not in agreement with the observed velocity profile (black line). {\it Panel (e):} Velocity profile fitted with two Gaussian components ($\chi^2$/doF=3.44). {\it Panel (f):}  Velocity profile fitted with three velocity components highlighted in blue for the brighter one, magenta for the blueshifted and cyan for the redshifted component. The dashed red line is the best fit  with a $\chi^2$/doF=0.54. Below each graph is a plot of the fit residuals.}
  \label{Fit_profile}
\end{figure*}

\subsection{Profile fitting} 
\label{Profile_fitting}
To obtain  the kinematic information of a specific region or object, we can extract the integrated line-of-sight ($LOS$) velocity profiles over a window that contains the whole object or region, in order to adjust the minimum number of Gaussian functions necessary to reproduce the observed profile. 
The velocity profiles were fitted by hand using the QtiPlot - Data Analysis and Scientific Visualisation Program\footnote{QtiPlot is a scientific application for data analysis and visualisation that uses least-squares algorithms for linear and nonlinear fitting of experimental data. \url{https://www.qtiplot.com/}}. 
The number of components fitted to each velocity profile was determined using a combination of visual inspection and the $\chi^2$/doF\footnote{The reduced chi-square ($\chi^2$/doF) is obtained by dividing the $\chi^2$ by the degrees of freedom (doF) and it is a method for model assessment and error estimation in astronomy.} statistic fit output by QtiPlot; this program performs a Gaussian fit with a single or multiple components depending on the velocity profile of the selected region or object to analyse.  In order to show the velocity profile fitting process, we present in Figure \ref{Fit_profile} three examples of  representative H$\alpha$ velocity profiles fitted for different objects or regions within NGC\,1569.
Below each graph we also show a plot of the fit residuals.

The observed velocity profile (black line) fitted with a single velocity component (red dashed line) is given in Figure \ref{Fit_profile}a, this region corresponds to HII region no. 12 \citep[of radius r= 5.6$\arcsec$,][]{Waller1991} integrated over a box of 11 px $\times$ 11 px (12.9$\arcsec \times$ 12.9$\arcsec$) which is a typical velocity profile of a HII region. In this case, the observed velocity profile is symmetric i.e., does not present wings or humps, therefore the profile was fitted with a single Gaussian.  The $\chi^2$/doF=0.25.

An example of an observed velocity profile fitted with two components is shown in Figures \ref{Fit_profile}b  and \ref{Fit_profile}c, which correspond to the profile of the pixel shown in the bottom-left panel of Figure \ref{Central} (see Section \ref{VF}). As a first step, we fitted a first component highlighted with the grey doted line ($\chi^2$/doF = 5.1, panel b). Visually, this fit did not match the observed velocity profile (black line); we then fitted it with two components highlighted in blue and magenta. The  best (or final) fit is shown in the red dashed line,  which presents a better visual fit, statistical parameters ($\chi^2$/doF=0.41) and better residuals than the single component fit.  In some cases a third Gaussian component was needed, as shown in Figure \ref{Fit_profile}d, which corresponds to the velocity profile of SNR16 (see Section \ref{SNR_param}). When fitted with a single component ($\chi^2$/doF= 8.87) is possible to appreciate small wings in its sides, indicating the presence of more than one velocity component. Then, we fitted two components which are highlighted in blue and magenta (see Figure \ref{Fit_profile}e). The final fit is shown in the red dashed line, with statistical parameters ($\chi^2$/doF=2.06). Finally, this velocity profile was fitted with three velocity components shown in blue for the brighter one, magenta for the blueshifted component and cyan for the redshifted component (see Figure  \ref{Fit_profile}f). The red dashed line shows the best fit  with a $\chi^2$/doF= 0.47, with better residuals than those obtained fitting one- and two-component.

It is noteworthy that given the natures and shell expanding morphology of SNR, the velocity profiles associated with this objects can be fitted with three components \citep[see e.g.][]{Sanchez-Cruces2018,Rosado2021}: the central and principal component related to the systemic velocity of the galaxy and the secondary components related to the expansion velocity. Therefore, as we showed, the SNR velocity profile was fitted with three Gaussian components as expected.
Also, the signal-to-noise ratio (S/N) of the velocity components of SNR16 profile are above the noise level (S/N$\gg$5) (see Table \ref{Kinematics_Parameters}).

The different morphological features of the velocity profiles show the nature of the ionised gas morphology  throughout the galaxy.
We noticed that the velocity profiles of the central parts of the galaxy (see Figure \ref{Central})   show a similarity, all of them were fitted with two velocity components, a broad and a narrow component. 
The narrow component is related to the stellar component, while the broad component is related to winds and stellar formation.  Therefore, we were able to spatially study the different morphology of the velocity profiles from the central part of the galaxy (see Section \ref{O-region}). On the other hand, the morphology of the SNRs velocity profiles, which are fitted with three components, is related to the systemic velocity of the galaxy represented by the main velocity component, and to the shell expansion morphology represented by the secondary components \citep[see][Figure 5]{Ambrocio-Cruz2004}.

\label{F3}
\begin{figure}\centering
\includegraphics[width=1\columnwidth]{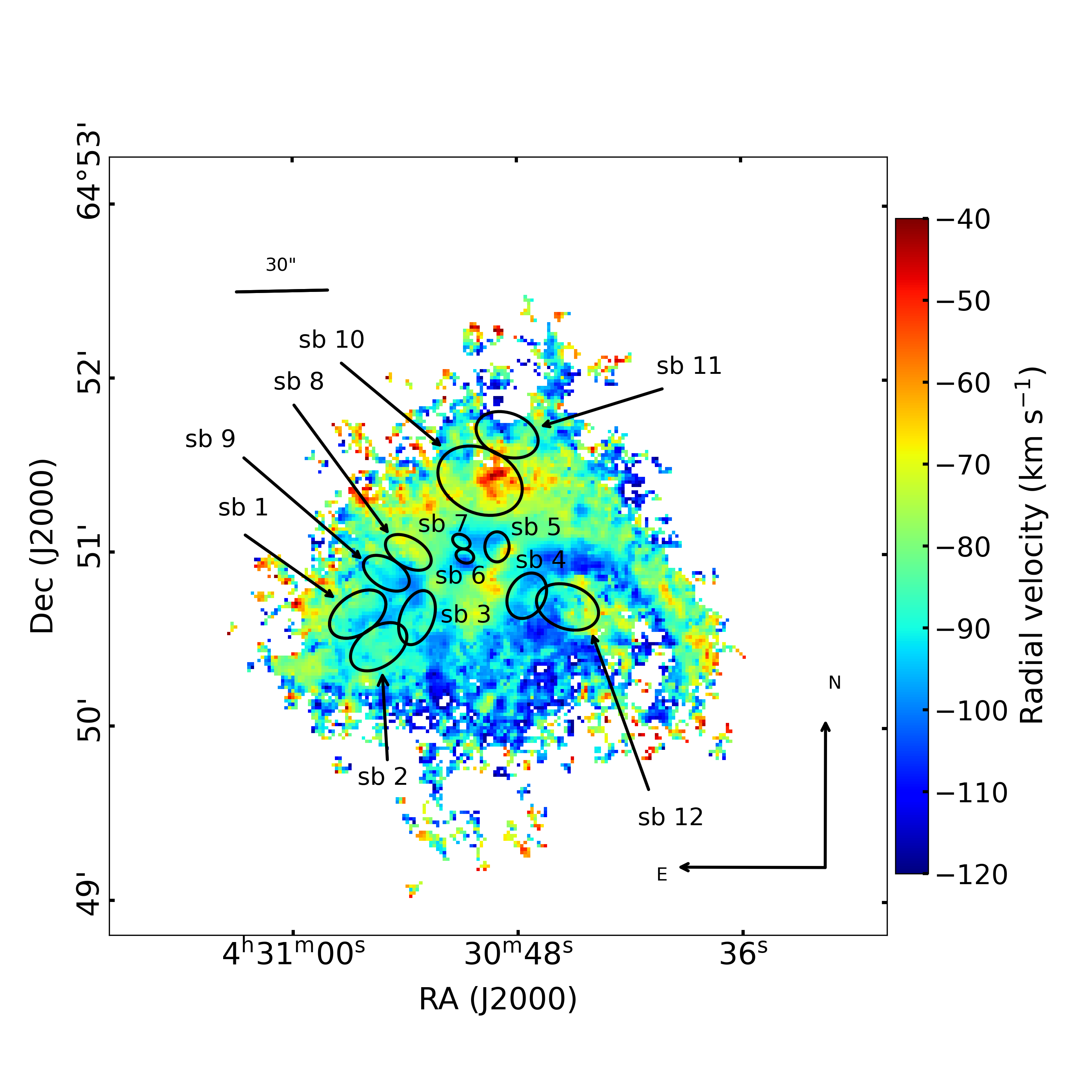}
  \caption{H$\alpha$ velocity field of NGC\,1569 obtained with PUMA. Position of superbubbles studied by \citet{Sanchez-Cruces2015} are overplotted.}
  \label{RV_Ha}
\end{figure}

\label{F4}
\begin{figure}\centering
\includegraphics[width=\columnwidth]{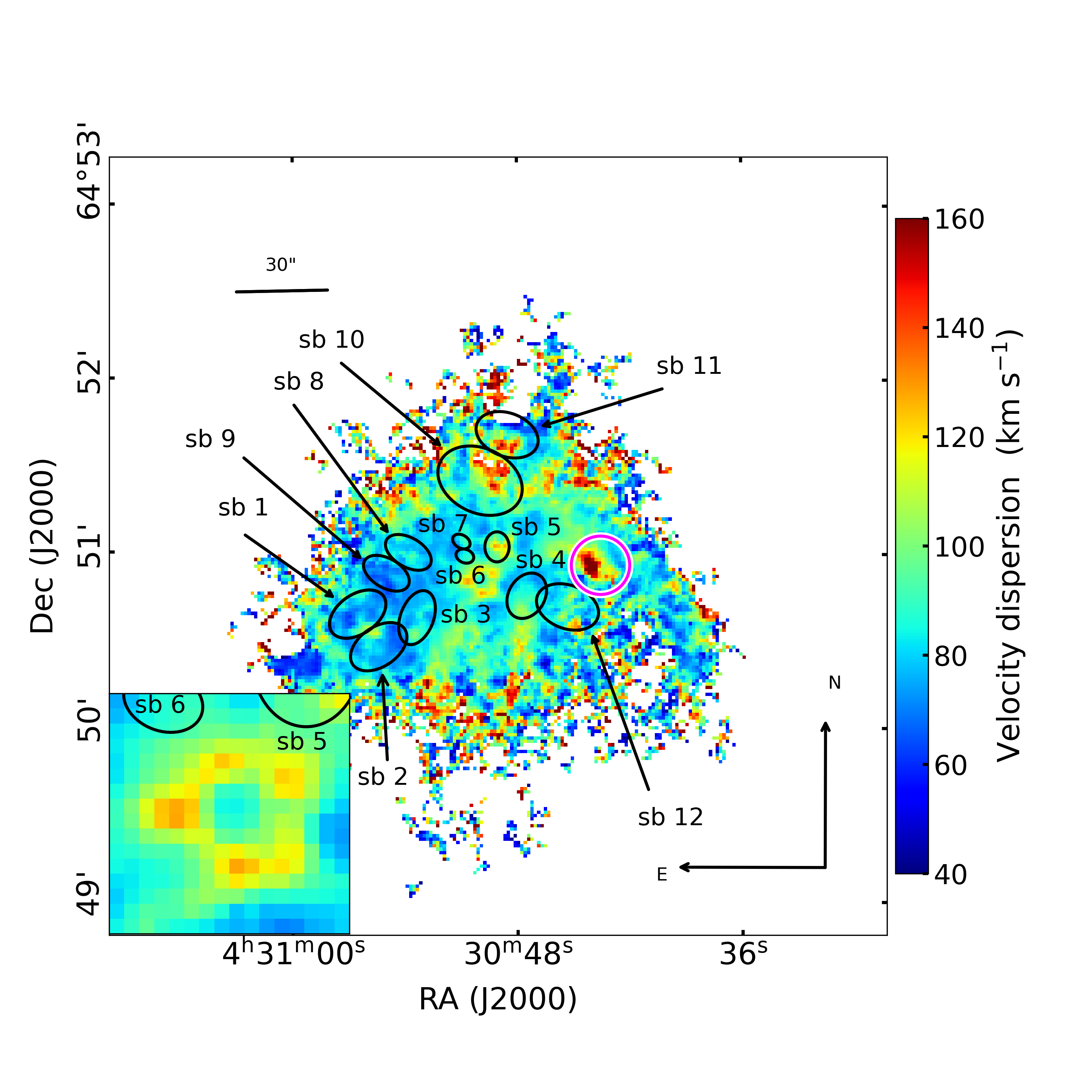}
  \caption{H$\alpha$ velocity dispersion map of NGC\,1569, The position of the superbubbles of Paper I are overplotted with ellipses. One of the highest velocity dispersion regions non related with any superbubble is highlighted with the pink circle. Inset panel shows a close-up of the central part of the galaxy showing other region with high velocity dispersion values non related with any superbubble. Inset panel size is 12$\times$12 pix (226.7 $\times$ 226.7 pc).}
  \label{FWHM_sb}
\end{figure}

\label{kin} 
\section{Kinematical analysis of NGC 1569}

\subsection{Velocity Fields}\label{VF}
Figure \ref{RV_Ha} displays the velocity field (VF) of NGC\,1569 computed from the H$\alpha$ emission line with the superbubbles of Paper I overplotted. 

The VF shows that the H$\alpha$ emitting gas does not seem to follow circular rotation, and that the gas motions are disordered due to the presence of the superbubbles and SNRs hosted in this galaxy. The VF displays velocity values ranging from -105 km s$^{-1}$ to -47 km s$^{-1}$ implying a difference of about $\Delta$V $\sim$60 km s$^{-1}$. This value is within the range of $\Delta$V values in nearby dwarf galaxies presented by \citet{Moiseev2014} using FP observations of ionised gas. On the northeast region of the galaxy, a high blueshifted region is seen with line-of-sight velocity $V_{LOS}$ = -46  km s$^{-1}$ which corresponds to the superbubble sb10 identified in Paper~I.

Figure \ref{FWHM_sb} shows the H$\alpha$ emission line velocity dispersion map of NGC\,1569 with the superbubbles classified in Paper~I overplotted. In general, the velocity dispersion values of NGC\,1569 are around 40-90~km~s$^{-1}$ except for five regions: one located on the western side of the galaxy  with velocity dispersion values of $\sim$165 km~s$^{-1}$ and pointed out with a pink circle in  Figure \ref{FWHM_sb}. This region matches the region where the arm of the galaxy comes out. The second, third  and fourth regions correspond to the location of superbubbles 5, 10 and 11, respectively, with velocities of about 118 km~s$^{-1}$ for the region related to the sb5, 143 km~s$^{-1}$ for the region related to sb10 and 150 km~s$^{-1}$ for the region related to the sb11. The fifth  region is located in the central part of the galaxy (see inset panel in Figure \ref{FWHM_sb}) and presents velocity dispersion values ranging between 118 and 127 km~s$^{-1}$. A velocity profiles inspection in this last region shows that the velocity profiles in the bottom-right and central parts present humps, while other velocity profiles seem to be single and broadened (see Figure \ref{Central}). In this same Figure, we highlighted with circles the nearly annular shape outlined by the high velocity dispersion values, which from now on will be call the `O' region.

This region is not related to any superbubble, but seems to be associated to the stellar cluster No. 35 classified by \citet{Hunter2000} located in the middle of the annular shape, and to the Super Stellar Cluster B \citep[SSCB,][]{Arp1985} located in the top part of the outer edge of the `O' region.

\subsubsection{Kinematics of the `O' region}
\label{O-region} 

In order to analyse the kinematics of the `O' region, we fitted the velocity profiles (as described in Section \ref{Profile_fitting}) of the pixels within a FoV of 15 $\times$ 15 pix (17.7 $\times$ 17.7\arcsec) encompassing this region.

\begin{figure*}\centering
\includegraphics[width=2\columnwidth]{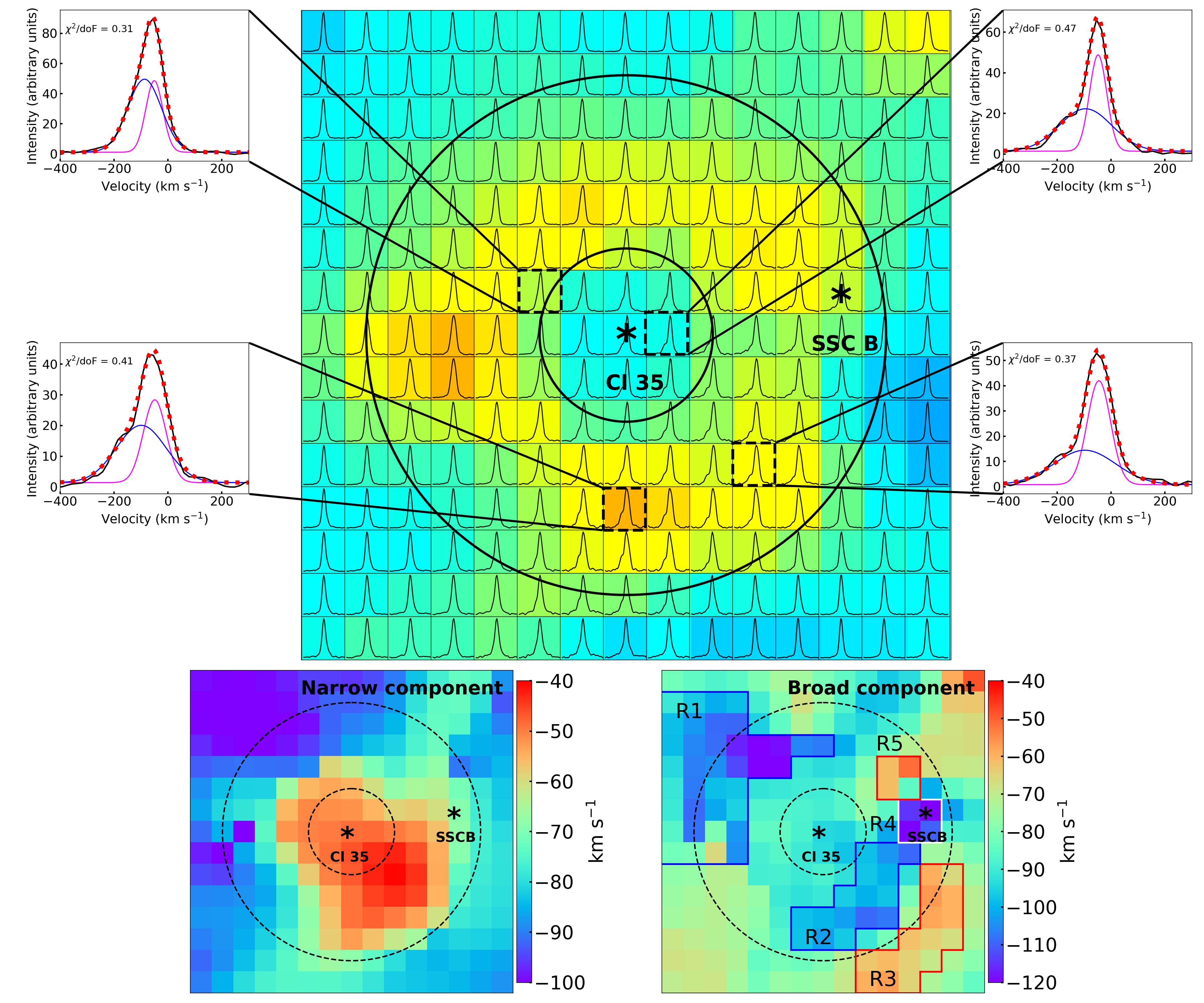}
  \caption{\textit{Top-panel}: Velocity dispersion map of the central part of the galaxy with velocity profiles  (in black) overplotted and surrounded by  a few representative velocity profiles fitted with narrow (in  pink) and broad (in blue) components used to build the velocity map of each velocity component.  Best fit is shown in red dashed line. Circles show the annular shape (the `O' region).  The position of the stellar cluster No. 35 and SSCB are indicated. \textit{Bottom-panel}: Velocity maps of the narrow (left) and broad (right) components of the central part of the galaxy. Dotted lines show the position of the annular shape of the `O' region. For all maps the position of the stellar cluster No. 35 and SSCB are indicated. R1 to R5 in broad velocity map are the representative regions described in Section \ref{O-region}.}
  \label{Central}
\end{figure*}

We found that all velocity profiles could be fitted with two velocity components, a narrow and a broad one, even those profiles that seemed to be single (for example top-left velocity profile in Figure \ref{Central}), therefore, all of them were fitted using these two components. The average values of the FWHM of the narrow and broad components are $\sim$72 $\pm 12$ km~s$^{-1}$ (in pink) and $\sim$184 $\pm 49$ km~s$^{-1}$ (in blue), respectively.
Using the barycenter method for both the broad and the narrow  velocity components, we obtained the respective velocity field of each component shown in the bottom panels of Figure \ref{Central}. Bottom-left panel shows the velocity map of the narrow component and bottom-right panel shows the velocity map of the broad component.

The velocity map of the narrow component presents velocity values of approximately -60 km~s$^{-1}$ to -40 km~s$^{-1}$ in the region surrounded by the annulus associated with the `O' region (see Figure \ref{Central}); velocity values for pixels located on the southwest (SW) side of the `O' region remain constant at about -80 km~s$^{-1}$, while for pixels located on the northeast (NE) side of the `O' region velocities remain constant with lower values of about -100 km~s$^{-1}$.
Interestingly, this VF shows a fan-like distribution of V$_{LOS}$ with values between -65 km~s$^{-1}$ and -40 km~s$^{-1}$ that seem to be centred near the location of stellar cluster No. 25  and ``opens up'' towards the SW side of the `O' region.
Following its location, this feature could be tracing one component of an expanding shell centred in stellar clusters No. 35.

As for the velocity map of the broad component, it presents five distinctive regions: {\it i)} Region R1 located on the NE of the `O' region has pixels with velocity values going from -110 to -100 km~s$^{-1}$. R1 seems to follow the external circumference of the `O' region.  {\it ii)} Region R2 located on the SW side within the `O' region where pixels present velocities between -110 and -100 km~s$^{-1}$. {\it iii)}  Region R3 located SW of the `O' region along R2 presenting velocities between -60 and -50 km~s$^{-1}$. {\it iv)} Region R4 located in the position of the SSCB with velocity values of -120 km~s$^{-1}$. {\it v)} Region R5 located  NE  of the SSCB inside the `O' region with velocities of about -50 km~s$^{-1}$.
This fifth region could actually be an extension of R2 with part of R3 superposed between R2 and R5. 

The structure of regions R1 and R2 might indicate the presence of an expanding shell of ionised gas produced by the stellar cluster No. 35 situated in the centre of the `O' region. This cluster is one of the most massive after SSCA, SSCB, and Cl 30. 
On the other hand, the high velocity values of region R4 might be related to the winds of the massive stars in SSCB. 

The behaviour in the `O' region is similar to that reported by \citet{Moiseev2012} using FP interferometry to study ionised gas velocity dispersion in dwarf galaxies. These authors identified and analysed supersonic turbulent motions and their relation with chaotic ionised gas motions and processes of current star formation.
They also found that high velocity dispersion values belong to the diffuse low surface brightness emission surrounding star-forming regions -see Figures 4, 5 and 6 in that work.
In the case of NGC\,1569, the `O' region is located where the monochromatic image displays rather low intensity values (right panel of Figure \ref{Ha_snr}) displaying a similar trend of the one mentioned by the authors above.
The presence of double components in the H$\alpha$ profiles in the central parts of dwarf irregular galaxies has also been studied by \citet{Egorov2021}. In their work, the analysis of the small-scale kinematics of the ionised gas indicated that the presence of composite profiles seemed to be associated with star-formation (SF) related phenomena (see their Figure 8).

\label{F5}

\subsection{Rotation Curve} 
\label{CR}
The rotation curve (RC) of NGC\,1569 has been computed in H$\alpha$  by \citet{Castles1991} and in several works from HI observations \citep[see for example][]{Reakes1980, Stil2002, Johnson2012, Lelli2014, Iorio2017}.

The H$\alpha$ RC from \citet{Castles1991} was derived using long-slit spectroscopy along the two brightest H$\alpha$ knots in the galaxy. Their RC shows almost constant rotation velocity along the galaxy disc displaying  a small rotation velocity difference of $\sim$10 km~s$^{-1}$  between positions -35 arcsec and +35 arcsec equivalent to -570 pc  and +570 pc considering a distance of 3.36 Mpc (see their Figure 3). They also used this Figure to emphasise the difficulty of measuring any meaningful rotation for this galaxy.

One of the most complete works regarding the determination of the RC of NGC\,1569 using HI observations was done by \citet{Iorio2017}  who studied the HI kinematics of 17 dwarf irregular galaxies extracted from the `Local Irregulars That Trace Luminosity Extremes, The HI Nearby Galaxy Survey' (LITTLE THINGS), using the $^{3D}$BAROLO software \citep{DiTeodoro2015}. \citet{Iorio2017} computed the RCs and the circular velocity (Vc) correcting for the asymmetric drift for all their sample, by fully exploiting the information in the HI data cubes to derive the RC of each galaxy using a modified tilted-ring method\footnote{3DB is a 3D method that performs a tilted-ring analysis on the data cube as a whole. \citet{Iorio2017} directly model the HI disc 3D data cube (two spatial axes plus one spectral axis) without explicitly extracting velocity fields.}. 

Using this approach, data cube models were convolved with the instrumental response, so the final results are not affected by the beam smearing.
Once the RC is obtained, these authors applied the asymmetric-drift correction in order to derive the Vc of each galaxy.  
Regarding NGC\,1569, \cite{Iorio2017} found the ISM of this galaxy to be highly turbulent with a velocity dispersion of $\sim$20 km s$^{-1}$ and an asymmetric-drift correction dominating at all observed radii (up to 2.5 kpc).
They explained the flattening of their HI RC solely to the asymmetric-drift correction (see their Figures 18 and 24).
These authors pointed out that the kinematic data reported for NGC\,1569 should be used with caution especially at the inner radii (from 0.0 kpc to 0.75 kpc) since no significant rotation of the gas is observed in this interval. 
According to these authors, their best-fitting model for NGC\,1569 is a good representation of the large-scale structure and kinematics of the HI disc (from 0.75 kpc outwards), though it fails to reproduce small-scale local features. 

\citet{Lelli2014} also found a similar result regarding the validity of the RC for the inner parts of NGC\,1569.

\subsubsection{RC computation}
\label{RC_computation}
As we show in Figure \ref{RV_Ha}, NGC\,1569 does not show a velocity gradient that  can be clearly associated with disc rotation which points in the direction of previous works mentioned above regarding the disturbed kinematics of this galaxy, especially in the inner parts.

However, we attempted nonetheless to derive a RC from the H$\alpha$ line VF of the galaxy following the methodology in \citet{amram1996,FuentesCarrera2004,Sardaneta2020} taking into consideration the detection of multiple velocity components in the inner parts of the galaxy; in particular, those found in the `O' region and discussed in Section \ref{O-region}.

The RC of the galaxy was obtained with the ADHOCw software using the standard tilted-ring method, in which the galaxy is described using a set of concentric elliptical rings. In this case, we considered a width of two-pixels (0.377~kpc), with their main axis, position angle (PA), systemic velocity $V_{sys}$, centre position $(x_0, y_0)$ and inclination ({\it i}) matching with those of the galaxy disc. 
The procedure to compute the RC with ADHOCw is done manually in an iterative manner where the initial parameters are manually modified in order to obtain a RC with low dispersion values and as symmetric as possible when the approaching and receding sides are superposed \citep[e.g.][]{amram1992,amram1996, FuentesCarrera2004,Repetto2010,Sardaneta2020}.
Usually in the case of isolated spiral galaxies, one of the requirements to derive the RC is for the inner parts on each side of the RC (receding and approaching velocities) to superpose, since the velocity fields are smooth and symmetrical resulting in symmetric and low-scattered rotation curves \citep{amram1996}.
This requirement can even be valid in the case of disc galaxies in early galaxy encounters \citep{FuentesCarrera2004}.

However, following the analysis in Section \ref{O-region} and previous results from HI observations \citep{Lelli2014,Iorio2017} discussed later in Section \ref{RC_1569}, we do not expect the assumption of RC symmetry to apply for the inner of parts of NGC\,1569.
In fact, an essential assumption for the previous method to be valid is that the gas moves in purely circular orbits, i.e. the contribution of radial and perpendicular velocities in the frame of the disc of the galaxy are negligible compared to the tangential velocity\footnote{In this frame of reference, the tangential velocity is usually called ``rotation velocity', $V_{rot}$}. With this in mind, one can describe the line-of-sight velocity, V$_{LOS}$ for any position $(x, y)$ on a ring with radius $R$ as:

\begin{equation}
V_{LOS} (x,y)= V_{sys}+V_{rot} (R) \cos \theta \sin i
\label{eq1}
\end{equation}

where, $\theta$ is the angle with respect to the receding major axis measured in the plane of the galaxy and $V_{rot}$ is the rotation velocity  \citep[see ][]{amram1996, FuentesCarrera2004,Sardaneta2020}.

This is clearly not the case for the inner parts of NGC\,1569. However, in order to derive the RC considering the presence of non-circular motions and following the analysis done in Section \ref{O-region}, we used a radii larger than 250 pc.

As initial parameters, we used the photometric parameters given by \cite{jarett2003} from the 2MASS\footnote{Two Micron All Sky Survey (2MASS).} Large Galaxy Atlas (LGA)\footnote{\url{https://irsa.ipac.caltech.edu/data/LGA/}}: 
the position of the photometric centre, $\alpha_{2MASS} (J2000)$= 04$^h$30$^m$49.18$^s$, $\delta_{2MASS}(J2000)$ = +64$\degr$50$\arcmin$47.78$\arcsec$; the position angle PA in the K-band, PA$_{2MASS}$ = 117.5$\degr$, and the axis ratio b/a=0.46.
We estimated the inclination ({\it i$_{2MASS}$=63.2$\degr$}) of the stellar galaxy disc assuming that the galaxy is an oblate spheroid using  $\cos^{2}i=((b/a)^{2}-p^{2})/(1-p^{2})$, where  $p=0.1$ is the intrinsic flattening of the galaxy \citep[e.g.][]{haynes1984}. The systemic velocity considered was V$_{sys}$=-104 ~km~s$^{-1}$ (NED).
We considered pixels inside a radius of 100 pix (1.3 kpc) and within an angular sector of 20$\degr$ along the galaxy major axis, making sure to only use pixels inside the galaxy disc and not those that could introduce dispersion (e.g., pixels in the galactic wind).

The values for the kinematical parameters were thus chosen to obtain a symmetric curve for regions of the galaxy where tangential velocity values are larger than radial and perpendicular velocities in the frame of the galaxy. The latter velocities can be partially assessed through the dispersion velocity values, i.e. the FWHM of the H$\alpha$ profile in each pixel -for more detail see \citet[][Section 5.2]{FuentesCarrera2004}.
Finally, to obtain the kinematic parameters that minimise the dispersion in the points along the RC beyond 250 pc up to 1.3 kpc, we made variations on the initial parameters.

Following the method described above, we derived a RC with the following  kinematical parameters: kinematical centre  $\alpha_{K}(J2000)$  = 04$^h$30$^m$49.36$^s$, $\delta_{K}(J2000)$ = +64$\degr$50$\arcmin$50.90$\arcsec$,  systemic velocity V$_{syst}$ = -88~km~s$^{-1}$ (value of the V$_{LOS}$ on the location of the kinematic centre), inclination \textit{i$_K$} = 61$\degr$ and position angle  PA$_K$ = 123$\degr$ (measured counterclockwise). Comparing the values of the H$\alpha$ kinematic parameters with the near-infrared photometric parameters from 2MASS, we notice they differ. The 2MASS photometric centre differs from the kinematic centre by 2.8$\arcsec$($\sim$2.74 pc); the PA$_{2MASS}$  differs from PA$_{K}$ by $\sim$5$\degr$ and {\it i$_{2MASS}$} differs from {\it i$_{K}$} by $\sim$2$\degr$. Finally, we obtained a systemic velocity within the range of the velocity reported in NED and the one reported by \citet{Iorio2017}. The derived RC is shown in the middle panel of Figure \ref{RC_1569}.

In the innermost region of the RC (radii smaller than 0.1 kpc), the rotation velocity values of the approaching and receding sides are separated by $\sim$40 $\mathrm{km\,s^{-1}}$, with the rotation velocity of the receding side reaching $\sim$20 $\mathrm{km\,s^{-1}}$ while the rotation velocity on the approaching side reaches $\sim$ -20 $\mathrm{km\,s^{-1}}$. 
After this radius (R=0.1 kpc), the difference in velocities between the receding and approaching side decreases until reaching a radius of 0.2 kpc when the RC becomes symmetric with a rotation velocity value close to 0 $\mathrm{km\,s^{-1}}$.
Between radii of 0.2 kpc and 0.75 kpc the RC stays symmetric showing an oscillating behaviour around a value slightly smaller than 0 $\mathrm{km\,s^{-1}}$.
Subsequently, between 0.8 kpc and 0.9 kpc, the receding and approaching side of the galaxy display rotation velocity differences of $\sim$15 $\mathrm{km\,s^{-1}}$. 
From 0.9 kpc to 1.0 kpc, the RC becomes symmetric again displaying an ascending behaviour with rotation velocity values going from 5 $\mathrm{km\,s^{-1}}$ to 10 $\mathrm{km\,s^{-1}}$.
For larger radii (R>1.0 kpc), the RC is no longer symmetric. 
At R=1.1 kpc, the rotation velocity difference reaches its largest value of $\sim$50 $\mathrm{km\,s^{-1}}$. 
From that radius up to the last H$\alpha$
emission point, the RC remains asymmetric (see Figure \ref{RC_1569}).

\subsubsection{VF and RC correction}
The VF considered to compute the RC in the previous Section was derived using the barycenter method mentioned in Section \ref{VF}. However, the detailed analysis of the `O' region in Section \ref{O-region} shows the presence of composite H$\alpha$ profile that seems to be tracing different kinematical phenomena in the innermost parts of NGC\,1569.
Considering that the broad component is probably related to winds from the stellar clusters present in the region, we decided to compute the RC of NGC\,1569 subtracting this broad component from the VF.
This was done by replacing the $V_{LOS}$ of the original VF with the velocity associated solely to the narrow component of the pixels displaying a composite profile -see Section \ref{O-region}. With this new VF, a new RC was derived following the methodology described in the previous Section. The corrected RC is shown in the bottom panel of Figure \ref{RC_1569}.

There is no significant difference between the kinematic parameters of the RC derived in Section \ref{RC_computation} that we shall call ``original RC'' and those of the RC derived in this Section that we shall call ``corrected RC''. 
The only parameter that differs is the kinematic centre located at $\alpha_{K-C}(J2000)$  = 04$^h$30$^m$49.36$^s$, $\delta_{K-C}(J2000)$ = +64$\degr$50$\arcmin$52.06$\arcsec$ for the ``corrected RC''.  Therefore, the ``corrected RC'' differs from the ``original RC'' by 1.16$\arcsec$ (18.9 pc). In the bottom-right inset panel of Figure \ref{RC_1569} we point out the 2MASS photometric centre ($\bigtriangleup$), the kinematic centre of this work (‘x’), the corrected kinematic centre of this work (`+') and that one reported by \citet{Iorio2017} ($\square$). The PA and sector used to determine the ``corrected RC'' is pointed out with a bold dashed line.

The middle and bottom panel of Figure \ref{RC_1569} allowed us to compare both RCs derived from the H$\alpha$ observations from PUMA.
Globally both RCs show a similar behaviour, with strong asymmetries in the innermost regions (R < 0.25 kpc).
However, for the ``corrected RC'', the velocity difference between both sides of this RC (approaching and receding sides) reaches a larger value of $\sim$ 65 km s$^{-1}$ for radii smaller than 0.09 kpc.
In spite of this difference, points between 0.02~kpc to 0.09~kpc for the ``corrected RC'' do not display the increasing/decreasing velocity  towards 0 km s$^{-1}$ trends for the approaching/receding sides of the RC seen in the ``original RC''.
Between 0.25~kpc to 0.5~kpc, both sides of the ``corrected RC'' show a symmetric behaviour displaying a larger slope than the slope of the ``original RC''. Nevertheless, values still lie close to the zero rotational velocity axis.
From 0.5~kpc to 0.75~kpc, the approaching and receding sides of the ``corrected RC'' show larger asymmetries than for the ``original RC,'' though globally (look at the star-like points in the bottom panel indicating the average rotation velocity for both sides of the RC) show a   similar behaviour: almost no rotation velocity is detected up to a radius of 0.75~kpc.
From 0.75~kpc to 1.1~kpc, the average RC for both H$\alpha$ RCs shows a similar behaviour displaying mostly negative values around -5 km s$^{-1}$.
From 1.1~kpc up to the last detected point in H$\alpha$, the ``corrected RC'' shows smaller dispersion than the ``original RC''.
For the last points detected, corresponding to the approaching side of NGC\,1569 rotation velocity, values lie between -10 km s$^{-1}$ to 7 km s$^{-1}$, while the ``original RC'' displays larger dispersion for the last detected points.

In summary, except for slight differences at different radii intervals, both the ``original RC'' and the ``corrected RC'' show a similar behaviour.
Neither RC shows rotation velocities smaller than -5 km s$^{-1}$ or larger than +10 km s$^{-1}$ for radii between 0.25~kpc and 0.75~kpc, nor do they display a smooth rotation velocity gradient between 0.75~kpc and 1.1~kpc.
Both of them reproduce the behaviour mentioned by \citet{Castles1991} regarding the H$\alpha$ RC derived with long-slit spectroscopy.

On the other hand, the fact that neither the RC could be derived through the matching of the approaching and receding sides in the innermost parts of the galaxy, along with the analysis of the composite profiles in those regions, nor the velocity dispersion analysis of the `O' region, indicates that indeed, the velocity dispersion of ionised gas in the central parts of NGC\,1569 reflect motions related with the gravitational potential of the galaxy. Moreover, the fact that neither RC shows a symmetric behaviour in the innermost parts of the galaxy, nor the rotation velocity exceeds an absolute value above 110 km s$^{-1}$; indicates that the velocity dispersion of the ionised gas in NGC\,1569 does not reflect virial motions in the gravitational potential of this galaxy.
Instead, the velocity dispersion is mainly determined by the energy injected into the ISM by ongoing SF as previously mentioned by \citet{Moiseev2015}.
These motions are also traced by the presence of composite emission line profiles analysed in Section \ref{O-region}.

In such a way that for NGC\,1569, we cannot derive a RC that truly traces the gravitational potential of the galaxy, since not only for R < 0.25 kpc, but for radii up to $\sim$ 1.1 kpc, the velocities plotted on the RCs do not represent rotation velocities inferred shown in equation \ref{eq1}.

\subsubsection{Comparison with HI observations}
Although the kinematic parameters given by \citet[][centre, i, PA]{Iorio2017} slightly differ from ours, it is possible to compare the kinematics between the atomic and ionised hydrogen due to the Iorio data resolution (15'').

\begin{figure*}\centering
\includegraphics[width=1.5\columnwidth]{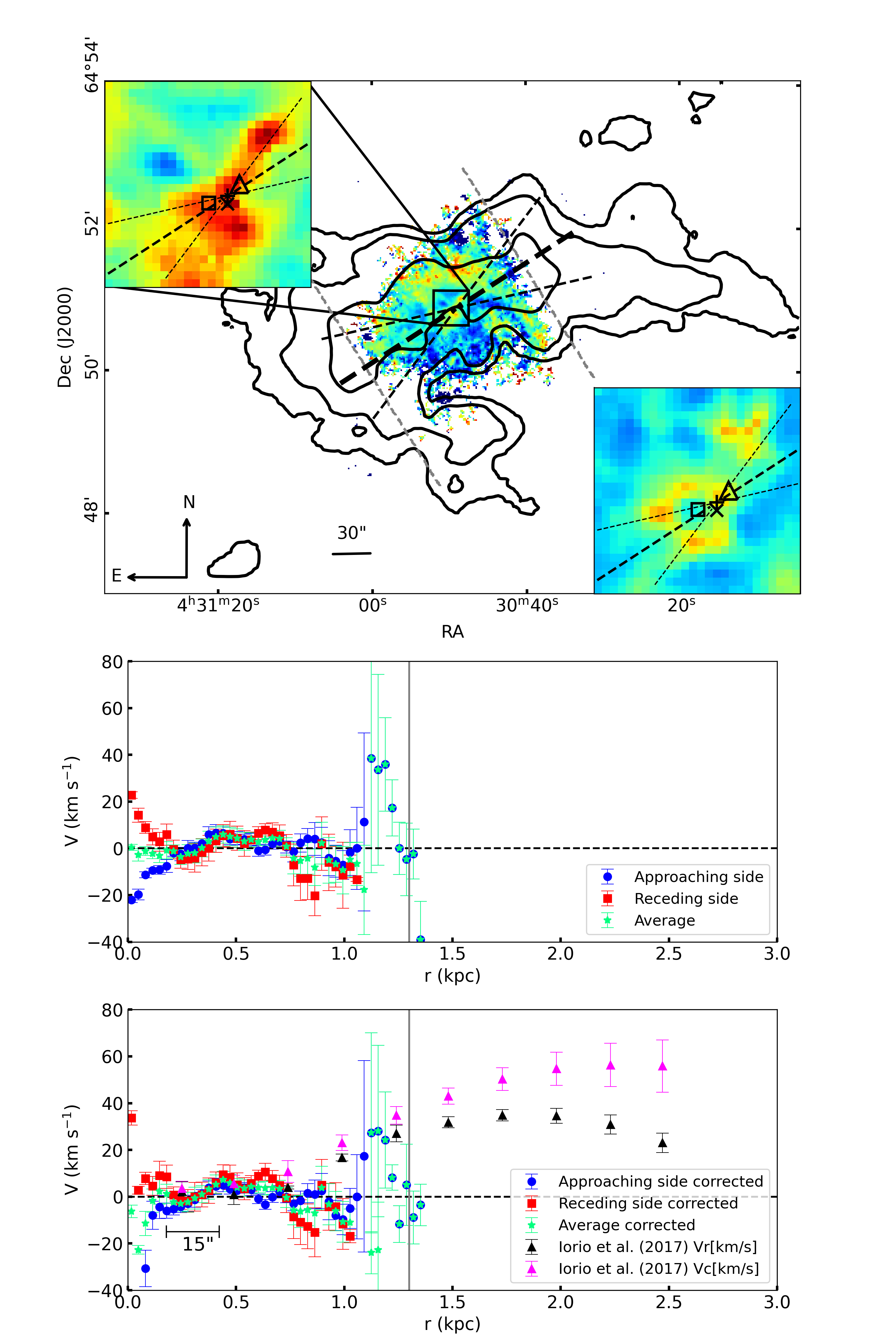}
  \caption{\textit{Top panel}: H$\alpha$ velocity field of NGC\,1569 with the angular sector used to analyse the H$\alpha$ rotation curve. Contours represent the HI emission from \citet{Iorio2017}. Top-left inset panel shows a close-up of the central region. Bottom-right inset panel shows the dispersion velocity of the central part of the galaxy. In both inset panels the angular sector used to compute the corrected H$\alpha$ rotation curve is shown as well as the 2MASS photometric centre ($\bigtriangleup$), the kinematic centre of this work (`x’), the corrected kinematic centre of this work (`+') and the one reported by \citet{Iorio2017} ($\square$). \textit{Middle panel}: The rotation curve of NGC\,1569 derived from the H$\alpha$ velocity field. \textit{Bottom panel}: The rotation curve of NGC\,1569 derived from the corrected H$\alpha$ velocity field (see text for more information). Pink and blue triangles correspond to the HI circular velocity (Vc) and best-fit rotation velocity (Vr) derived by \citet{Iorio2017} with a resolution of 15$\arcsec$ pointing out.  For middle and bottom panels, the receding side corresponds (red squares) to the southeastern side of the galaxy and the approaching side (blue circles) to the northwestern side of the galaxy. With green stars we show the average of the receding and approaching sides. The vertical grey line represents the 1.3 kpc radius used in this work to determine the optical RCs.}
\label{RC_1569}
\end{figure*}

Top panel of Figure \ref{RC_1569} shows the H$\alpha$ VF of NGC\,1569 obtained with PUMA with the HI contours of the galaxy from \citet{Iorio2017} overplotted. We can see the region covered by the H$\alpha$ observations in comparison to the HI emissions.
In the top-left inset panel of Figure \ref{RC_1569} we point out the 2MASS photometric centre ($\bigtriangleup$), the kinematic centre of this work (‘x’), the corrected kinematic centre of this work (`+') and the one reported by \citet{Iorio2017} ($\square$). The separation between our kinematical centre and the one obtained by \citet{Iorio2017} is about 2.9$\arcsec$ ($\sim$47.2 pc).

As previously mentioned, ionised gas motions in NGC\,1569 do not trace the gravitational well in the central parts of the galaxy. This is confirmed by the comparison between the H$\alpha$ RCs derived from our PUMA observations with the HI RC derived by \citet{Iorio2017}, shown in the bottom panel of Figure \ref{RC_1569}.
Regarding the part of the HI RC that these authors identified as non-representative of circular velocity (R < 0.75 kpc), we can see that the H$\alpha$ RCs show no definite velocity gradient on either side (approaching or receding), but instead trace the presence of winds associated with star cluster No. 35 and SSC B through our two-dimensional analysis.
We can thus say that the absence of circular motions detected by \citet{Iorio2017} and shown by their asymmetric-drift correction of the rotational velocity of this galaxy are traced in detail thanks to the high spectral and spatial resolutions of our PUMA observations.

\label{F6}

\subsection{Rotation Sense of NGC 1569}

Spiral arms can be classified by their orientation relative to the direction of rotation of the galaxy. A trailing arm is one whose outer tip points in the opposite direction to the galactic rotation, while the outer tip of a leading arm points in the direction of the rotation \citep[see,][]{binney-tremaine-2008}. 

Transitory one-armed leading spirals can be produced by plausible dynamical processes, for example, encounters with companion galaxies on retrograde orbits \citep[e.g.][]{grouchy-2008}.
 
To determine whether the arm of a
given galaxy leads or trails, it is necessary to figure out which side of the galaxy is closer to us using both kinematic and photometric data \citep{Pasha1985, FuentesCarrera2004, Repetto2010b,Sardaneta2020}. 

In the case of NGC\,1569, the H$\alpha$ direct image in the left panel of Figure \ref{Ha_snr} traces an arm-shaped feature to the west called the “Western  arm” by \cite{hodge-1974}. It has been suggested that the arm is associated with both outflow \citep[][]{waller-1991, martin-1998} and inflow \citep{Stil2002} episodes. Additionally, \cite{ott-2005} proposed that it is a H$\alpha$ tail emitting soft X-rays. Finally,  \cite{Johnson2013} proposed a scenario in which the Western arm is the product of two dwarfs merging. This last scenario might be possible considering that NGC\,1569 is associated with the IC 342 galaxy group.

\begin{figure}\centering
\includegraphics[width=\columnwidth]{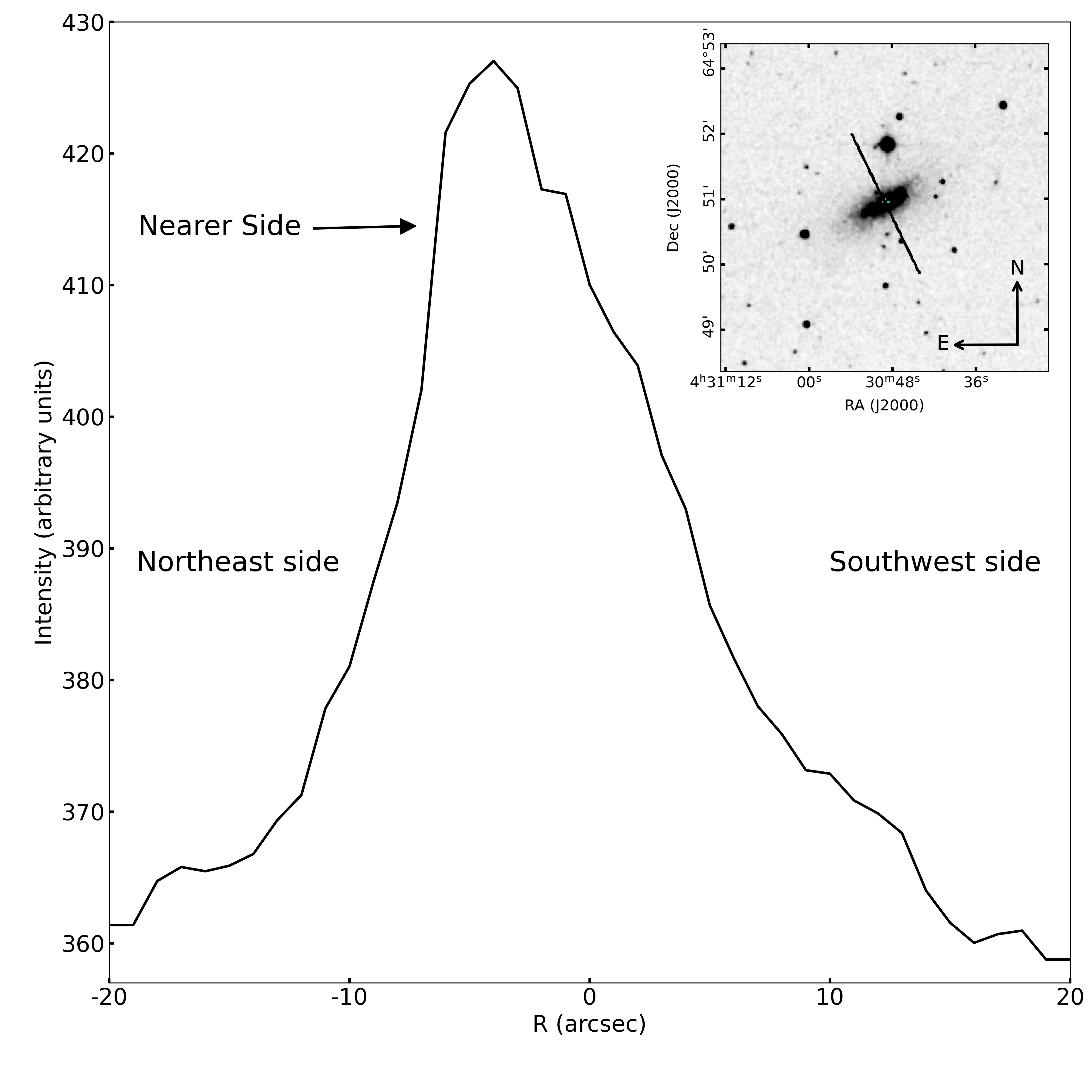}
  \caption{Intensity profile along the kinematic minor axis of NGC\,1569. The northern side of the profile falls more abruptly indicating this side is near. Top-right inset panel shows the K-band image from 2MASS and the cross section taken to obtain the profile.} 
  \label{rot_sence_1569}
\end{figure}

\begin{figure*}\centering
\includegraphics[width=2\columnwidth]{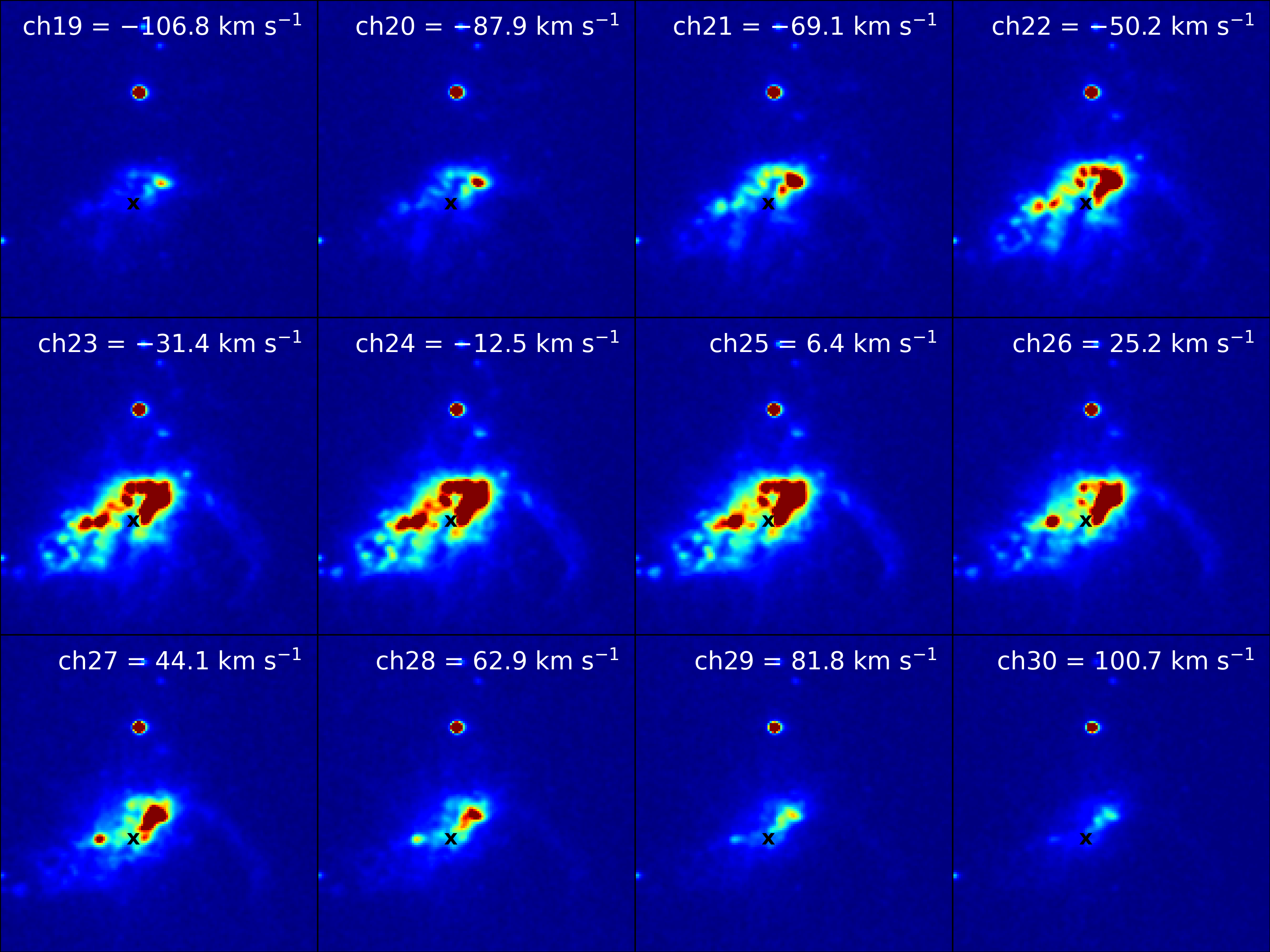}
  \caption{Characteristic channels maps of NGC\,1569 with continuum subtracted. The velocity of each channel is given with respect to the galaxy systemic velocity (-88$\pm$1 km s$^{-1}$). The `x' represents the kinematical centre.}
  \label{mosaic_1569}
\end{figure*}

Considering the merger scenario proposed by \cite{Johnson2013}, we verified the rotation sense of the Western arm. First, our kinematic information: the rotation curve in Figure \ref{RC_1569} and the H$\alpha$ $LOS$ velocity maps displayed in Figure \ref{mosaic_1569}, shows the receding $LOS$ velocities are at the northwestern side of the galaxy and that the approaching $LOS$ velocities are at the southeast. Second, to determine which side of the galaxy is closer to us we used the gradient criterion of \citet{Pasha1985}: because the dust obscures a fraction of the bulge light, the apparent brightness of the nuclear region falls off asymmetrically from the centre outwards along the minor axis, the far side being the one where the surface-brightness profile declines more smoothly. We obtained the surface-brightness profile of the nuclear region along the minor axis of NGC\,1569 (see Figure \ref{rot_sence_1569}) using the K-band image from 2MASS \citep{jarett2003}, concluding that the northeastern side is the nearest because the profile falls off more abruptly than along the southern side. 
Therefore, as a result of the photometric analysis to the K-band image and the kinematics of the galaxy disc, we conclude that the Western arm has leading rotation; i.e., the tip of the Western arm points to the same direction than the galactic rotation.

On the other hand, the isocontours of the HI emission in the top panel of Figure \ref{RC_1569} display two arm-like features,  called western and eastern butterfly wings by \citet{Johnson2012} and \citet{Johnson2013}; both of them pointing to the southwest. Using the rotation curve from \cite{Iorio2017} and our photometric analysis to the infrared image of NGC\,1569, it can be determined that the western arm (western butterfly wing) has leading rotation while the eastern arm (eastern butterfly wing) is trailing. 

Theoretical studies \citep{Athanassoula1978} and n-body simulations \citep{Thomasson1989, Byrd2008, Lieb2022} have shown that the encounter of two galaxies with opposite rotational directions (retrograde encounter) can generate spirals with a single long-lived leading arm (m = 1) if the retrograde companion is sufficiently massive (strong perturbation). It has also been proposed that leading material can be the result either from a dwarf dwarf merger \citep{Braun1992, Jore1996, Starkenburg2019} or from the infall of material with counter-rotation onto the galaxy \citep{Jore1996, Lovelace1998}. The size of NGC\,1569 and the morphology of its arms traced by the HI emission supports the hypothesis of two dwarf merger scenario.

\label{F7}

\label{F8}  

\section{Kinematical analysis of SNRs}
As we mentioned in Section \ref{SNR}, NGC\,1569 harbours 31 SNRs located in the galaxy disc. In this Section we studied the kinematics of ionised gas around their position and their physical parameters.

\subsection{Expansion Velocity of SNRs}
\label{exp_vel}
In order to determine the expansion velocity of the SNRs, we extract the H$\alpha$ velocity profiles of the ionised gas around their positions, integrated in windows of 3~pix $\times$ 3~pix (56.6\,pc $\times$ 56.6\,pc) due to the radius of the SNRs being $\lesssim$40 pc (see Table \ref{List_SNR}). The velocity profiles are complex; i.e., they present evident wings or humps and were fitted with three Gaussian functions as described in Section \ref{Profile_fitting} (see Figure \ref{SNR-ind-profile}). The main component (in blue) is associated with the rotation velocity of the gas around the centre of the galaxy. The secondary components (in magenta and cyan) are related to the expansion velocity, the best ﬁt is
shown in a red dotted line. The expansion velocities were obtained considering the expansive  motion of a shell and assuming that the optical emission is not located at the centre of the supernova remnant, therefore the expansion velocity is $V_{\rm exp}$ = ($V_{\rm max}$-$V_{\rm min}$) where $V_{\rm max}$ and $V_{\rm min}$ are the blueshift and redshift components, respectively. In order to estimate the quality of the fit of the velocity components, we took into account the $\chi^2$/doF value (pointed out in each velocity profile) and the S/N ratio of the different velocity components.

Table \ref{Kinematics_Parameters} shows the velocity values for each Gaussian component as well as their respective S/N ratio and the SNRs expansion velocity. 
As we see in Table \ref{Kinematics_Parameters} and Figure \ref{SNR-ind-profile}, all main components present a high S/N ratio (S/N$\gg$3). We can see that the S/N ratio of the secondary velocity components generally show an acceptable S/N ratio (S/N>3), except those related to SNR13 which have a S/N$<$3. Even though the S/N of the Gaussian velocity components increases for the velocity profiles integrated over regions of several pixels (as is the case), it was not possible to obtain secondary velocity components for SNR13 with a good  S/N ratio. Then, this velocity profile cannot be considered in the analysis.
Regarding the expansion velocities, almost all SNRs present $V_{\rm exp}$ $>$100 $\pm$1 km~s$^{-1}$ except SNRs 3, 7 and 8 with $V_{\rm exp}$ of 95 $\pm$1, 90 $\pm$1 and 87 $\pm$1 km~s$^{-1}$, respectively. The $V_{\rm exp}$ value of SNR5 ($V_{\rm exp}$ = 114 $\pm$1 km~s$^{-1}$) is in agreement with the $V_{\rm exp}$ value reported by \citet{Ercan2018}.

\begin{table*}\centering
  \setlength{\tabcolsep}{1\tabcolsep}
  \caption{Velocity and signal-to-noise ratio of the velocity components of the SNRs line profiles in NGC1569.}
  \label{Kinematics_Parameters}
  \label{Table2}
  \scalebox{1}{
 \begin{tabular}{lccccccccccccccc}
    \hline
ID &
\multicolumn{1}{c}{V}&
\multicolumn{1}{c}{$V_{\rm min}$}&
\multicolumn{1}{c}{$V_{\rm max}$}&
\multicolumn{1}{c}{$V_{\rm exp}$}&
\multicolumn{3}{c}{Component intensity}&
\multicolumn{1}{c}{Noise} &
\multicolumn{1}{c}{S/N$_1$}&
\multicolumn{1}{c}{S/N$_2$}&
\multicolumn{1}{c}{S/N$_3$}\\
&
\multicolumn{1}{c}{(km s$^{-1}$)}&
\multicolumn{1}{c}{(km s$^{-1}$)}&
\multicolumn{1}{c}{(km s$^{-1}$)}&
\multicolumn{1}{c}{(km s$^{-1}$)}&
\multicolumn{1}{c}{C$_1$}&
\multicolumn{1}{c}{C$_2$}&
\multicolumn{1}{c}{C$_3$}&
\multicolumn{1}{c}{level}\\
\hline
SNR1	&	-84.2	&	-150.7	&	-20.6	&	130.0	&	303.14	&	41.16	&	28.23	&	8.33	&	36.4	&	4.9	&	3.4	\\
SNR2	&	-95.6	&	-130.7	&	-25.6	&	105.0	&	178.06	&	24.74	&	42.63	&	3.01	&	59.2	&	8.2	&	14.2	\\
SNR3	&	-84.2	&	-120.7	&	-25.6	&	95.0	&	241.83	&	39.92	&	28.23	&	6.40	&	37.8	&	6.2	&	4.4	\\
SNR4	&	-80.2	&	-130.7	&	-10.6	&	120.0	&	234.47	&	41.16	&	39.97	&	9.74	&	24.1	&	4.2	&	4.1	\\
SNR5	&	-80.6	&	-130.7	&	-16.6	&	114.0	&	162.12	&	28.84	&	42.63	&	9.18	&	17.7	&	3.1	&	4.6	\\
SNR6	&	-98.1	&	-150.7	&	-30.6	&	120.0	&	160.90	&	24.74	&	42.63	&	4.20	&	38.3	&	5.9	&	10.1	\\
SNR7	&	-98.1	&	-130.7	&	-40.6	&	90.0	&	147.41	&	24.74	&	42.63	&	4.33	&	34.0	&	5.7	&	9.8	\\
SNR8	&	-90.6	&	-125.7	&	-38.6	&	87.0	&	160.90	&	24.74	&	42.63	&	4.38	&	36.7	&	5.6	&	9.7	\\
SNR9	&	-85.6	&	-150.7	&	-8.6	&	142.0	&	45.64	&	5.74	&	7.19	&	1.17	&	38.9	&	4.9	&	6.1	\\
SNR10	&	-90.6	&	-160.7	&	-12.6	&	148.0	&	61.58	&	8.53	&	8.78	&	2.15	&	28.7	&	4.0	&	4.1	\\
SNR11	&	-98.6	&	-160.7	&	-12.6	&	148.0	&	99.59	&	8.53	&	8.78	&	1.98	&	50.3	&	4.3	&	4.4	\\
SNR12	&	-92.6	&	-150.7	&	-20.6	&	130.0	&	48.13	&	8.65	&	7.19	&	0.67	&	71.6	&	12.9	&	10.7	\\
SNR13	&	-92.6	&	-170.7	&	1.6	&	172.3	&	21.44	&	2.50	&	3.99	&	1.35	&	15.9	&	1.9	&	3.0	\\
SNR14	&	-84.5	&	-185.1	&	1.8	&	186.9	&	331.07	&	18.12	&	32.00	&	1.85	&	179.0	&	9.8	&	17.3	\\
SNR15	&	-86.4	&	-150.1	&	-10.2	&	139.9	&	638.62	&	61.62	&	64.75	&	1.72	&	372.3	&	35.9	&	37.8	\\
SNR16	&	-91.1	&	-185.2	&	3.5	&	188.7	&	90.32	&	15.26	&	6.67	&	0.49	&	182.8	&	30.9	&	13.5	\\
SNR17	&	-86.5	&	-165.9	&	-4.6	&	161.4	&	496.10	&	28.41	&	35.19	&	1.23	&	404.7	&	23.2	&	28.7	\\
SNR18	&	-93.6	&	-140.8	&	-35.2	&	105.6	&	50.85	&	12.96	&	7.67	&	1.09	&	46.8	&	11.9	&	7.1	\\
SNR19	&	-79.6	&	-149.7	&	-15.6	&	134.0	&	34.24	&	5.60	&	6.45	&	0.77	&	44.3	&	7.2	&	8.3	\\
SNR20	&	-79.6	&	-129.7	&	-19.6	&	110.0	&	30.63	&	5.60	&	7.62	&	1.07	&	28.7	&	5.2	&	7.1	\\
SNR21	&	-83.7	&	-149.7	&	-20.6	&	129.0	&	140.53	&	15.32	&	19.05	&	1.19	&	117.9	&	12.9	&	16.0	\\
SNR22	&	-89.7	&	-170.7	&	-25.6	&	145.0	&	90.57	&	18.00	&	4.88	&	0.76	&	119.5	&	23.7	&	6.4	\\
SNR23	&	-81.7	&	-137.1	&	-10.7	&	126.4	&	181.54	&	32.86	&	21.66	&	1.13	&	160.0	&	29.0	&	19.1	\\
SNR24	&	-99.7	&	-169.3	&	-55.2	&	114.1	&	27.87	&	5.60	&	6.45	&	0.89	&	31.4	&	6.3	&	7.3	\\
SNR25	&	-82.5	&	-137.7	&	-25.6	&	112.1	&	249.65	&	32.76	&	32.65	&	0.73	&	341.9	&	44.9	&	44.7	\\
SNR26	&	-86.4	&	-130.8	&	-15.8	&	115.0	&	75.01	&	11.99	&	8.14	&	1.36	&	55.2	&	8.8	&	6.0	\\
SNR27	&	-91.8	&	-150.7	&	10.5	&	161.3	&	57.81	&	8.43	&	5.36	&	1.07	&	54.1	&	7.9	&	5.0	\\
SNR28	&	-91.9	&	-160.2	&	-20.7	&	139.5	&	121.53	&	10.95	&	14.99	&	0.92	&	132.0	&	11.9	&	16.3	\\
SNR29	&	-100.5	&	-165.4	&	-45.4	&	120.0	&	167.13	&	16.32	&	14.96	&	0.82	&	205.0	&	20.0	&	18.4	\\
SNR30	&	-94.9	&	-155.1	&	-20.2	&	134.8	&	115.44	&	11.80	&	12.64	&	0.64	&	179.0	&	18.3	&	19.6	\\
SNR31	&	-96.9	&	-155.2	&	-18.9	&	136.3	&	113.67	&	8.48	&	5.39	&	1.15	&	99.1	&	7.4	&	4.7	\\
\hline
\end{tabular}}\\
\begin{flushleft}
The columns are as follows:\\
Column 1 : SNR identification\\
Column 2 : Main velocity component obtained from the fitted profiles. \\
Column 3-4 : Blueshift and redshift components,  ($V_{\rm min}$ and $V_{\rm max}$, respectively) from the fitted profiles .\\
Column 5 : Expansion velocity measured as $V_{\rm exp}$ = ($V_{\rm max}$-$V_{\rm min}$).\\
Column 6 : Intensity value of the main velocity component (C$_1$).\\
Column 7-8: Intensity value of  blueshift and redshift components (C$_2$ and C$_3$, respectively). \\
Column 9: S/N ratios of the main velocity component (S/N$_1$).    \\
Column 10-12: S/N ratios of the different blueshift and redshift components (S/N$_2$ and S/N$_3$, respectively).\\
\end{flushleft}
\end{table*}


\subsection{Physical Parameters of SNRs} 
\label{SNR_param}
By using Fabry-Pérot data in the H$\alpha$ and [{S\,{\sc ii}}]$\lambda\lambda$6717,6731 lines, we can derive the physical parameters of SNRs such as the  electron density ($n_{\rm e}$), kinetic energy ($E_{\rm k}$), and kinematic age ($t$) considering that the SNRs are in a determined evolutionary stage.
According to \citet{Woltjer1972}, the theoretical SNRs evolution is described by four phases: I) the free expansion. II) the adiabatic or Sedov-Taylor.  III) the radiative or the ``snow-plough''. IV) the merging phase. 
Then, to determine the evolutionary phase where the SNRs are, we considered the SNR evolution diagram which shows the relation between the blast-wave velocity, age  and radius for the four evolutionary stages \citep[see][]{Cioffi1990,Padmanabhan2001,Micelotta2018}.  

In summary, the diagram shows that when the explosion occurs, the  SNR expands at approximately constant velocity and  its radius, r, is proportional to t (r$\propto$t). When the remnant enters the adiabatic phase, its radius is generally described by the expression r$\propto$t$^{2/5}$, velocities reach thousands of km s$^{-1}$ and the diameters are between  $\sim$5 and $\sim$30 pc. When the remnant enters in radiative phase its radius is generally described by the expression r$\propto$t$^{1/4}$, the velocity decreases considerably down to about 200 km s$^{-1}$ and radius between $\sim$20 and $\sim$40 pc \citep[see][]{Berkhuijsen1987, Reid2015, Bozzetto2017}.

Then, considering the SNRs expansion velocity found in this work (from 87 to 188 km s$^{-1}$) and their sizes reported in the literature (between 21 to 37 pc), we concluded that the SNRs are in the radiative phase. 
Therefore, we were able to compute the physical parameters for the SNRs in NGC\,1569 considering that the SNRs are in the radiative phase.

\subsubsection{Electron density}
From the [{S\,{\sc ii}}] wavelength data cube, where both $\lambda$6717 \AA\ and $\lambda$6731 \AA\ lines were detected, we were able to compute pixel per pixel the [{S\,{\sc ii}}]$\lambda$6717/[{S\,{\sc ii}}]$\lambda$6731 line-ratio. In both cases we did not correct for differential extinction because those lines are too close in wavelength and any correction would be cancelled out when estimating ratio. 

To determine $n_{\rm e}$, we used the [{S\,{\sc ii}}]$\lambda$6717/[{S\,{\sc ii}}]$\lambda$6731 ratio of the SNRs sample; these line-ratio values span from 1.08 $\pm$0.02 to 1.33 $\pm$0.03. Under the assumption that S$^{+}$ emission is formed in a region between 5000 K and 10$^{4}$ K and using  the \textit{temden} routine of STSDAS/IRAF\footnote{`Image Reduction and Analysis Facility' \url{http://iraf.noao.edu/}. IRAF is distributed by National Optical Astronomy Observatory, operated by the Association of Universities for Research in Astronomic, Inc., under cooperative agreement with the National Science Foundation.}, we calculated $n_{\rm e}$ assuming an electron temperature $T_{\rm e}$ = 10$^{4}$ K (as an order of magnitude estimated for any nebula with near-solar abundances, \citet{Osterbrock1989}). Thus, we obtained a range of electron density between 90 $\pm$29 and 432 $\pm$32 cm$^{-3}$ (see Table \ref{Physical_Parameters}).

If we consider that the shock is radiative, where the compressed gas is cooled by collisions, the pre-shock electron density $n_{\rm 0}$ is given by:

\begin{equation}
\label{eq2}
n_{0} = n_e\left(\frac{c_s}{V_s}\right)^2
\end{equation}

where $c_{\rm s}$ = 10~km~s$^{-1}$ is the sound speed of the environment at $T_{\rm e}$ = 10$^{4}$~K and $V_{\rm s}$ = $V_{\rm exp}$ is the shock velocity in km~s$^{-1}$. The pre-shock densities of SNRs rank from 0.5 $\pm$32 to 3.9 $\pm$32 cm$^{-3}$. The errors of $n_{0}$ were computed applying the error propagation to equation \ref{eq2}.\footnote{ The error of n$_0$ is: $\delta n_{0} = n_{0} \sqrt{\left(\frac{\delta n_e}{n_e}\right)^2 +  \left(2\frac{\delta V}{V}\right)^2}$ where $\delta n_e$ and $\delta V$ are the error of the electron density and velocity, respectively.}

\subsubsection{The Energy and Age of the SNRs}
We can determine the age and energy of the SNRs considering that they are in the radiative phase with the  numerical model of \citet{Chevalier1974}:

\begin{equation}
\label{eq3}
t(4) =   30.7 R/V_s 
\end{equation}

\begin{equation}
\label{eq4}
E_{50} = 5.3\times10^{-7}n_0^{1.12}V_s^{1.4}R^{3.12}  
\end{equation}

where $V_{\rm s}$ = $V_{\rm exp}$ is the shock velocity in km~s$^{-1}$, R is the linear radius in pc, t(4) is the age of the remnant in units of 10$^{4}$~yr and $E_{\rm 50}$ is in units of 10$^{50}$~erg. The linear radii were taken from the literature from \cite{Chomiuk2009}; for SNRs without radius information (SNRs 8, 10 and 11) the average value was used (see Table \ref{List_SNR}). The errors of t(4) and $E_{50}$ were computed applying the error propagation to equations \ref{eq3} and \ref{eq4} \footnote{The errors of t(4) and E$_{50}$ are:} 
\blfootnote{$\delta t(4) = t(4) \sqrt{\left(\frac{\delta V}{V}\right)^2 + \left(\frac{\delta R}{R}\right)^2}$}
\blfootnote{$\delta E_{50} = E_{50} \sqrt{1.12\left(\frac{\delta n_0}{n_o}\right)^2 + \left(1.4\frac{\delta V_s}{V_S}\right)^2+ \left(3.12\frac{\delta R}{R}\right)^2}$}, respectively.

The values of the energy deposited in the ISM by the SN explosion obtained in this work are in the order of 10$^{50}$~erg, typical values of SNRs (see Table \ref{Physical_Parameters}).
The age of the SNRs found in this work considering that they are in the radiative phase is about $\sim$10$^{4}$ yr (see Table \ref{Physical_Parameters}). We were not able to obtain the age of SNR13 due to the poor detection of the [{S\,{\sc ii}}]$\lambda$6717/[{S\,{\sc ii}}]$\lambda$6731 ratio in this location.
Also, as we presented in the previous section, the S/N ratio of the secondary velocity components is too low to be considered. Therefore, we consider that it is not possible to study the physical properties of this SNR.

\begin{table}\centering
  \setlength{\tabcolsep}{1\tabcolsep}
  \caption{Physical Parameters of Supernova remnants of NGC 1569.} 
  \label{Physical_Parameters}
  \label{Table3}
  \scalebox{1}{
 \begin{tabular}{lccccccccccccccc}
    \hline
ID &
\multicolumn{1}{c}{n$_e$}&
\multicolumn{1}{c}{t(4)}&
\multicolumn{1}{c}{E(50)}\\
&
\multicolumn{1}{c}{(cm$^{-3}$)}&
\multicolumn{1}{c}{(10$^{4}$ yr)}&
\multicolumn{1}{c}{(10$^{50}$ erg s$^{-1}$)}\\
\hline
SNR1	&	142.2	$\pm$22	&	4.0	$\pm$0.06	&	2.8	$\pm$0.6	\\
SNR2	&	311.9	$\pm$46	&	5.5	$\pm$0.10	&	10.8	$\pm$0.7	\\
SNR3	&	153.4	$\pm$11	&	7.6	$\pm$0.16	&	10.8	$\pm$0.7	\\
SNR4	&	200.9	$\pm$24	&	3.1	$\pm$0.05	&	1.5	$\pm$0.2	\\
SNR5	&	213.5	$\pm$64	&	5.9	$\pm$0.10	&	10.8	$\pm$2.3	\\
SNR6	&	226.5	$\pm$13	&	3.7	$\pm$0.06	&	3.1	$\pm$0.2	\\
SNR7	&	213.5	$\pm$64	&	8.9	$\pm$0.20	&	22.3	$\pm$3.0	\\
SNR8	&	296.7	$\pm$30	&	6.0	$\pm$0.14	&	8.9	$\pm$0.4	\\
SNR9	&	281.9	$\pm$103	&	6.0	$\pm$0.08	&	25.2	$\pm$7.4	\\
SNR10	&	131.3	$\pm$43	&	3.5	$\pm$0.05	&	2.3	$\pm$1.4	\\
SNR11	&	239.8	$\pm$81	&	3.5	$\pm$0.05	&	4.5	$\pm$1.6	\\
SNR12	&	213.5	$\pm$103	&	4.0	$\pm$0.06	&	4.4	$\pm$1.8	\\
SNR13	&	---		            &	---	            &	---		\\
SNR14	&	176.5	$\pm$83	&		2.8$\pm$0.03	&		2.6$\pm$1.6	\\
SNR15	&	432.1	$\pm$38	&		3.8$\pm$0.05	&	9.1	$\pm$0.5	\\
SNR16	&	311.9	$\pm$77	&	2.1	$\pm$0.02	&	2.1	$\pm$0.7	\\
SNR17	&	267.5	$\pm$14	&	2.3	$\pm$0.03	&	1.7	$\pm$0.1	\\
SNR18	&	90.2	$\pm$29	&	5.0	$\pm$0.09	&	2	$\pm$0.9	\\
SNR19	&	142.2	$\pm$77	&	9.1	$\pm$0.14	&	38.4	$\pm$29.4	\\
SNR20	&	142.2	$\pm$33	&	3.4	$\pm$0.06	&	1.1	$\pm$0.2	\\
SNR21	&	153.4	$\pm$33	&	3.1	$\pm$0.05	&	1.3	$\pm$0.4	\\
SNR22	&	131.3	$\pm$32	&	2.9	$\pm$0.04	&	1.2	$\pm$0.6	\\
SNR23	&	110.3	$\pm$51	&	3.2	$\pm$0.05	&	0.9	$\pm$0.7	\\
SNR24	&	343.8	$\pm$32	&	5.0	$\pm$0.09	&	11.2	$\pm$0.6	\\
SNR25	&	239.8	$\pm$26	&	3.8	$\pm$0.07	&	3	$\pm$0.2	\\
SNR26	&	90.2	$\pm$29	&	3.9	$\pm$0.07	&	1.2	$\pm$0.6	\\
SNR27	&	131.3	$\pm$64	&	2.6	$\pm$0.03	&	1.1	$\pm$1.2	\\
SNR28	&	413.3	$\pm$92	&	3.4	$\pm$0.05	&	6.4	$\pm$0.8	\\
SNR29	&	267.5	$\pm$57	&	4.4	$\pm$0.07	&	6.1	$\pm$0.8	\\
SNR30	&	110.3	$\pm$41	&	3.2	$\pm$0.05	&	1.1	$\pm$0.7	\\
SNR31	&	153.4	$\pm$56	&	5.9	$\pm$0.09	&	10.9	$\pm$5.7	\\
\hline
\end{tabular}}\\
\begin{flushleft}
The columns are as follows:\\
Column 1 : SNR identification\\
Column 2 : Electron density calculated from the [{S\,{\sc ii}}]$\lambda$6717/[{S\,{\sc ii}}]$\lambda$6731 ratio.\\
Column 3 : Dynamical age computed using the numerical model of \citet{Chevalier1974}. \\
Column 4 :  Energy deposited in the ISM by the SN explosion computed using the numerical model of \citet{Chevalier1974}.\\
\end{flushleft}
\end{table}

\subsection{Surface Brightness - Diameter ($\Sigma$-D) Relationship}

Given that we are studying the full sample of SNRs in NGC\,1569, and that those SNRs are at the same distance, it is worth to search for a radio $\Sigma$-D relation between them.  This relation was first proposed by  \citet{Shklovskii1960} who derived it from theoretical grounds concerning the evolution of the SNRs radio-emission. The possible existence of such relation has important  implications mainly for the Galactic SNRs because it implies a method of distance determination in a similar way as the method, also proposed by Shkloivski, to estimate the distance to optically thin planetary nebulae. Also, it is an important relation  not only statistically but  because it gives evolutionary properties of SNRs samples. Statistically, this relation shows the differences of the radio surface brightness ($\Sigma$) and diameter (D) slope in different galaxies \citep[see ][]{Berkhuijsen1986, Fenech2008, Long2017} and therefore, the dependence of this relationship. Regarding the evolutionary properties; this relation can give information about the evolution of the SNRs sample such as the SN type, and supernova progenitor, total kinetic energy of the SN, the SNR evolution stage \citep{Urosevic2020} and  the ISM density. Then, all these parameters are different for every SN and the ISM where it evolves \citep{Kostic2016}.

Following the first theoretical derivation by \citet{Shklovskii1960}, the $\Sigma$-D relation is often described as a power law of the form:
\begin{equation}
\Sigma = AD^{-\beta}
\label{eq5}
\end{equation}

where parameter A and slope $\beta$ are obtained by fitting the observational data for a sample of SNRs. In particular, parameter A depends on the properties of both the SN
explosion and the ISM while $\beta$ is thought to be independent of these properties
\citep{Arbutina2005}.

This relation is also used for determining distances to  Galactic and extragalactic SNRs  \citep[i.e.,][]{Arbutina2004, Arbutina2005, Urosevic2005, Urosevic2010, Pavlovic2013, Pavlovic2014, Bozzetto2017, Vukotic2019}.  

For a better understanding of the density distribution of the local ISM in NGC\,1569 we plotted  in Figure \ref{sigma-d} the radio surface brightness of the 29 SNRs in function of their size by using the flux densities definition:

\begin{equation}
\Sigma= 1.505\times10^{-19}\dfrac{S}{\theta^2}\ \ [Wm^{-2} Hz^{-1} sr^{-1}] 
\label{eq6}
\end{equation}

where S is the integrated flux density in Jansky (Jy) and $\theta$ is the diameter in arcmin. The diameters and S values were taken from \citet{Greve2002} and \citet{Chomiuk2009}. In this case, those authors presented only the uncertainties for the S parameter which are very low (on average $\sim$0.09 mJy) to be taken into account.

By using QTiplot, we use a  linear fit slope to compute the dependency between this parameters, and obtained the relation:

\begin{equation}
\Sigma(D) = 3.36\times10^{-15} \times D^{-1.26\pm0.2} W m^{-2} Hz^{-1} sr^{-1}.
\label{eq7}
\end{equation}

Figure \ref{sigma-d} shows the $\Sigma$-D of the SNRs; as we can see, it presents significant scattering and high dispersion. We observed that $\Sigma$ decreases  as the diameter increases. The power-law slope ($\beta$) value obtained in this work  is  $\beta$ = 1.26 $\pm$0.2 which is comparable (within 2-3 sigma only) with the value obtained for SNRs in M31 ($\beta$ = 1.67) and M33 ($\beta$ = 1.77) in \citet{Urosevic2005} using a sample of 30 and 51 radio SNRs, respectively. However, our $\beta$ value is considerably different to other
galaxies with higher $\beta$, such as the sample of 40 radio SNRs in the LMC \citep[$\beta\sim$3.78,][]{Bozzetto2017}; the sample of 19 radio SNRs in the SMC \citep[$\beta\sim$5.16,][]{Filipovic2005, Bozzetto2017} and the sample of 31 radio SNRs in M82 \citep[$\beta\sim$3.9,][]{Urosevic2010}.

According to the \citet{Duric-Seaquist1986} theory  of the $\Sigma$-D relation for SNRs, the youngest SNRs should have $\beta \approx$5 and the oldest should have $\beta \approx$3.5; the SNRs in M82, SMC and LMC are in the Sedov-Taylor phase \citep[see,][]{Bozzetto2017, Duric-Seaquist1986, Berezhko-Volk2004}). Regarding the SNRs with $\beta \approx$2, \citet{Urosevic2005} found that the $\Sigma$-D is a statistical effect rather than a physical explanation. Therefore, due to our value, $\beta$<2, we did not get a different solution compared to that from the $\Sigma$-D relation one ($\beta$=2) trivial statistical one.

Considering the low density environment of NGC\,1569, it is expected that its SNRs have lower surface brightness than SNRs in dense environments, due to the dependence between surface brightness and density \citep[$\Sigma$, $\rho^{\eta}$, n$_H^{\eta}$,][]{Duric-Seaquist1986,Berezhko-Volk2004}. 
Therefore, as it is shown in the literature, there is a $\Sigma$–D relation for each extragalactic SNR sample that depends on the different SNR nature and environment of the host galaxy, see for example \citet{Urosevic2005} and  \citet{Bozzetto2017}, where a considerable range of $\beta$ values are shown (from $\beta \sim$2.58 to $\beta \sim$5.1).

\label{F9}
\begin{figure*}\centering
\subfloat{\includegraphics[width=2.2\columnwidth]{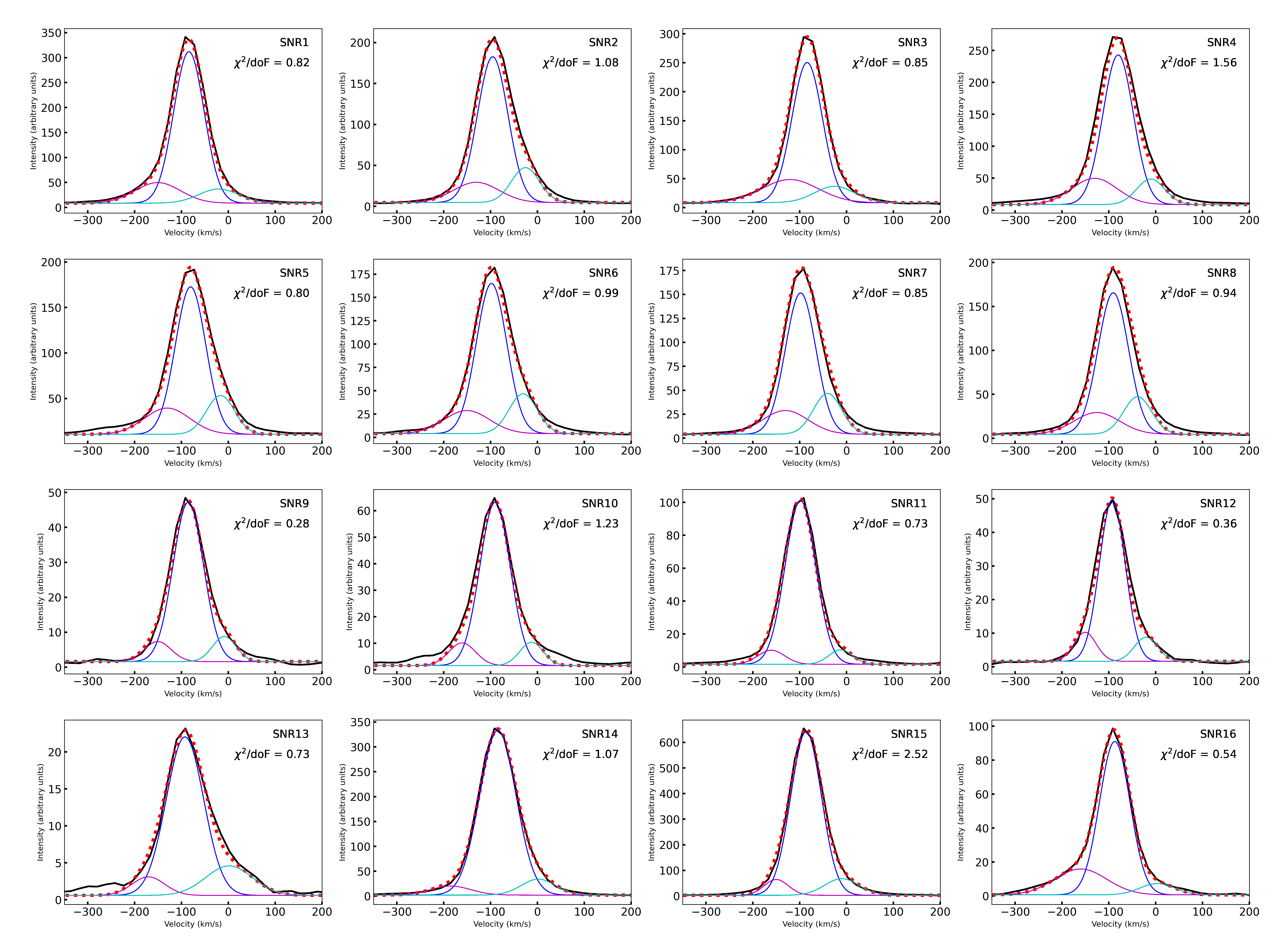}}
\caption{Fitted SNRs velocity profiles. For all of them, the original proﬁles is in black line. The blue velocity component refers to the main velocity component, the secondary velocity components are in pink and cyan. Red doted line represents the best fit.}
\label{SNR-ind-profile}
\end{figure*}

\label{F9continued}
\begin{figure*}\centering
\ContinuedFloat
\subfloat{\includegraphics[width=2.2\columnwidth]{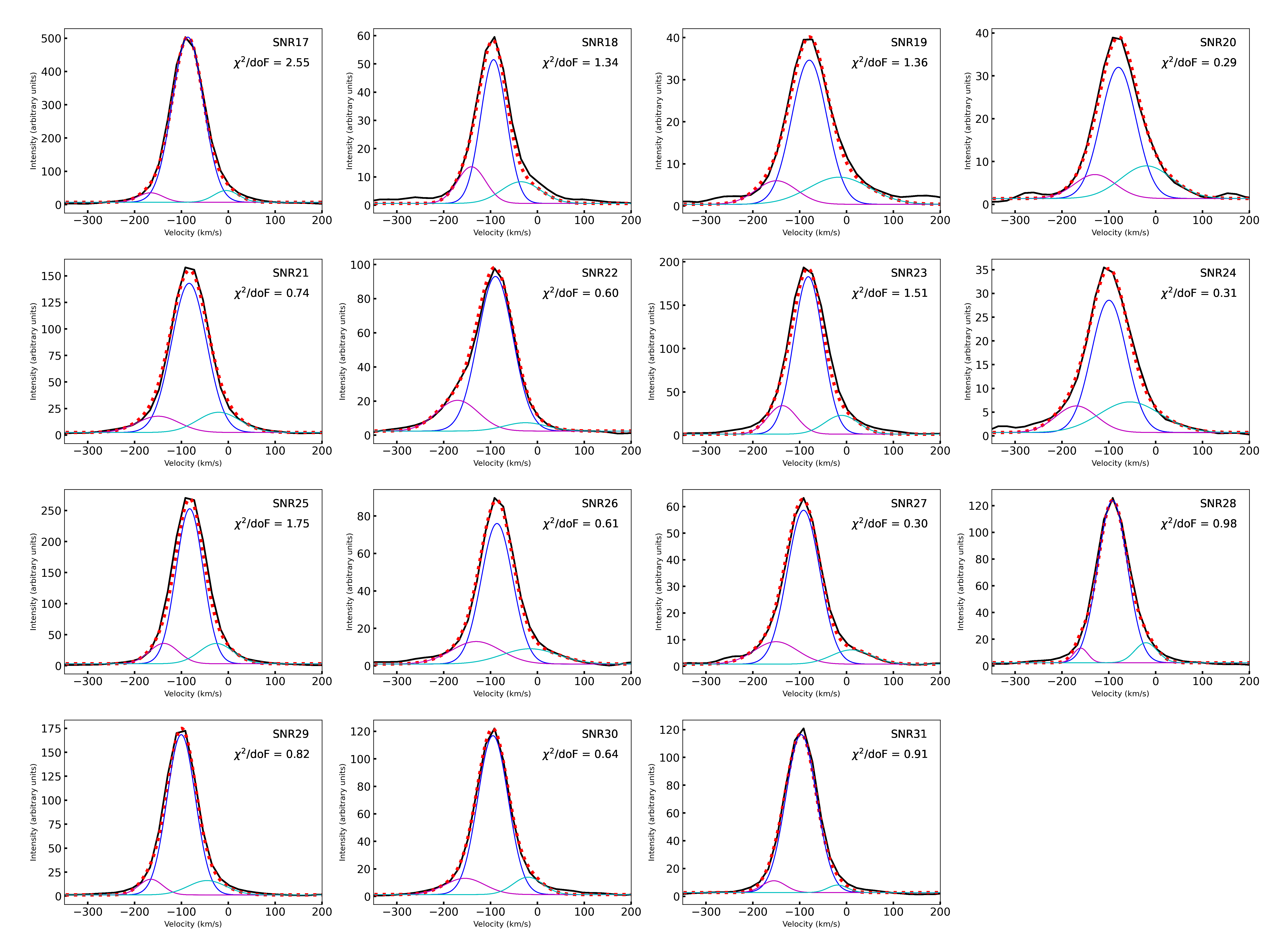}}
\caption{Continued}
\label{SNR-ind-profile}
\end{figure*}

\label{F10}
\begin{figure}
\includegraphics[width=1\columnwidth]{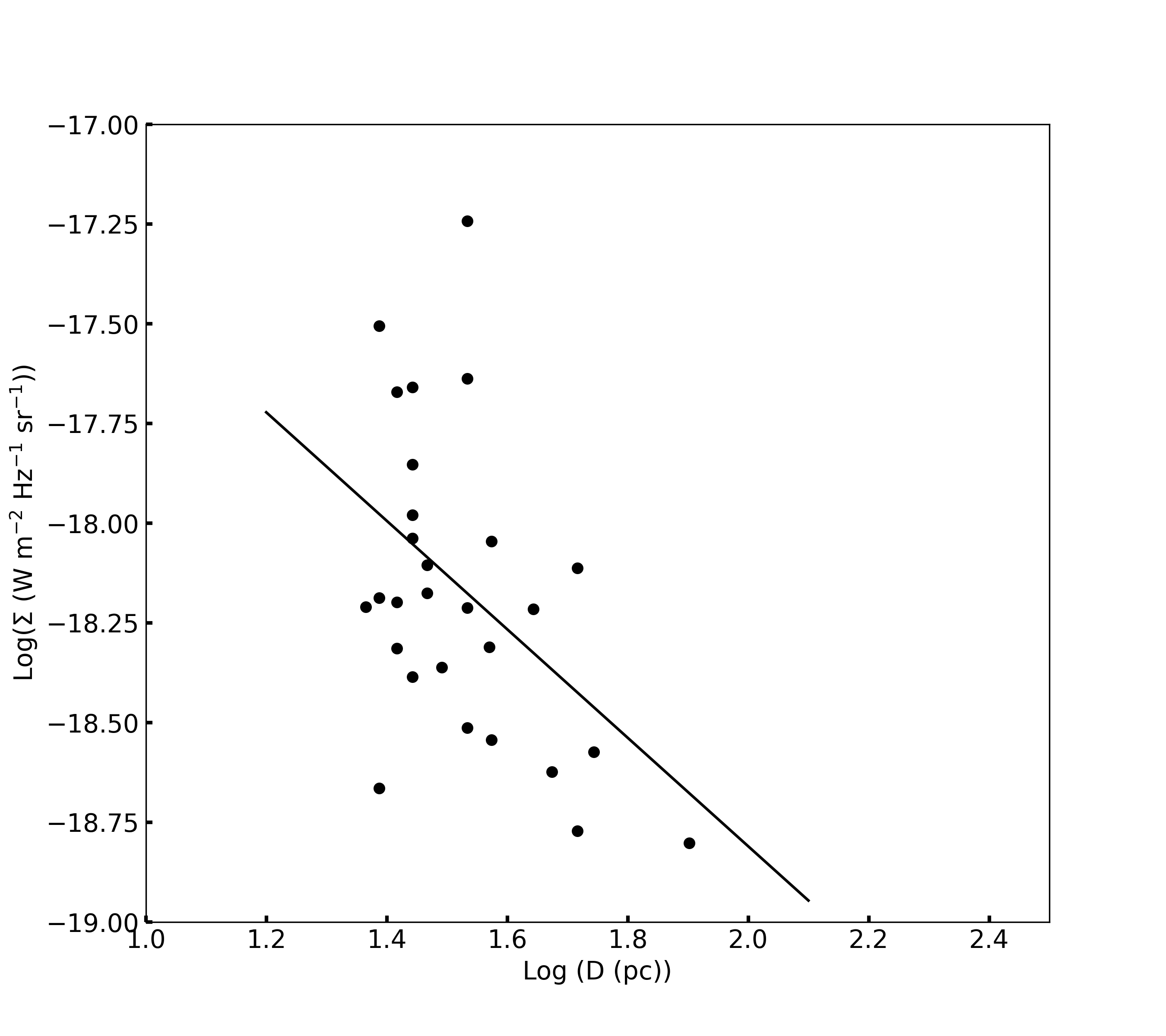}
\caption{The $\Sigma$-D relation for NGC\,1569 SNRs. Line represents the best fit given by: $\Sigma$(D)$_{H\alpha}$ =
3.36$\times$10$^{-15}$ $\times$ D$^{-1.26\pm0.2}$ W m$^{-2}$ Hz$^{-1}$ sr$^{-1}$.  Solid line is the least-squares fit.}  
\label{sigma-d}
\end{figure}

\section{Conclusions}
\label{conclusions}
In this paper, we studied the global kinematics of the dwarf irregular galaxy NGC\,1569 in the H$\alpha$ line and analysed the kinematics of the gas surrounding its hosted SNRs. The main results are summarised as follows.
  
The H$\alpha$ emitting gas of  NGC\,1569 does not show regular rotation in its velocity field and presents similar behaviour to other dwarf galaxies \citep[see][]{Moiseev2014}. 
The velocity dispersion map shows very interesting kinematical information of the central part of the galaxy. In particular, we found a region with high velocity dispersion we named the `O' region, which presents an annulus shape morphology and is related to stellar cluster No. 35 of \cite{Hunter2000}, the fourth most massive stellar cluster after stellar cluster No. 30, SSCA and SSCB.

The velocity analysis of this region might indicate on one hand, the presence of an expanding shell of ionised gas produced by the stellar cluster No. 35, and on the other hand, might be related to winds of massive stars coming from SSCB. In conclusion, the ionised gas in this region seems to be associated with SF related phenomena, as pointed out by \cite{Egorov2021}.
Also, this analysis of the `O' region confirms that the velocity dispersion of ionised gas in central parts of NGC\,1569 does not reflect the motions related with its gravitational potential.

The RCs we computed show the same behaviour than the RC obtained by \cite{Castles1991} with long-slit spectroscopy. Therefore, considering this result and the velocity dispersion analysis of the `O' region,  we were not able to derive a RC that truly traces the gravitational potential of the galaxy.
However, given the high spectral and spatial resolutions of our PUMA observations, it was possible to trace the ionised gas in the central parts of  NGC\,1569, as well as the absence of circular motions detected previously by \cite{Iorio2017} using HI observations. The disturbed kinematics of this galaxy could be related to the fact that this galaxy has a lower angular momentum, $j_*$ (given its  stellar mass, $M_*$) than the average angular momentum of the galaxies with comparable stellar mass \citep[see][]{Mancera-Pina2021}.

From the kinematic and  photometric analysis of this galaxy, we confirm that the Western arm points on the same direction of the galactic rotation, while the Eastern arm points on the opposite direction. This means that NGC\,1569 is an extraordinary case of a galaxy, with an arm leading rotation probably as a result of a two dwarf merging scenario.

Using our high spectral resolution observations, we determined the expansion velocities of 31 SNRs hosted in this galaxy. The velocity profiles of the SNRs are complex, indicating the presence of non circular motions. Almost all SNRs have a $V_{\rm exp}>$100 km~s$^{-1}$ indicating the presence of shocks. We also calculated the energy deposited in the ISM by the SN explosion and the age of 30 SNRs considering they are in the radiative phase. The age interval is 2.3 to 8.9 $\times$10$^{4}$ yr and the energy ranges from 1 to 39 $\times$10$^{50}$ erg s$^{-1}$, assuming all of them are in the radiative phase.

Finally, we calculated the $\Sigma$–D relation for SNRs in NGC\,1569 and obtained $\beta$ = 1.26$\pm$0.2 comparable to  the value determinednfor SNRs in M31 ($\beta$ = 1.67) and M33 ($\beta$ = 1.77), but not with other galaxies that have higher $\beta$ values, such as LMC ($\beta$ = 3.78), SMC ($\beta$ = 5.16) and M82 ($\beta$ = 3.9).  Nevertheless, our result must be taken with caution, since $\beta \sim$2 due to a statistical effect which produces the canonical $\Sigma\propto$D$^{-2}$ relation (since the definition of surface brightness is implied in that relation). Even though the slope we obtained does not have a physical meaning, we presented this relation to complement this work.

\section*{Acknowledgements}
We thank Dr. Iorio G. for providing us with the HI contours.
The authors acknowledge the financial support from DGAPA-PAPIIT (UNAM) IN109919, CONACYT CY-253085 and  CONACYT CF-86367 grants. I.F.C. acknowledges the financial support of SAPPI-20222041 grant. 
Based upon observations carried out at the Observatorio Astron\'omico Nacional on the Sierra San Pedro M\'artir (OAN-SPM), Baja California, M\'exico.
We thank the daytime and night support staff at the OAN-SPM for facilitating and helping obtain our observations.
Facilities: OAN-SPM, M\'exico
\section*{Data Availability}

The data underlying this article will be shared on reasonable request to the corresponding author.

\label{refecrencias}
\bibliographystyle{mnras}
\bibliography{references1569.bib} 

\label{lastpage}

\end{document}